\documentclass[11pt,english]{article}
\usepackage{mathpazo}
\usepackage[T1]{fontenc}
\usepackage[latin9]{inputenc}
\usepackage{geometry}
\usepackage{babel}
\usepackage{verbatim}
\usepackage{float}
\usepackage{rotfloat}
\usepackage{bm}
\usepackage{amsmath}
\usepackage{amsthm}
\usepackage{amssymb}
\usepackage{graphicx}
\usepackage{setspace}
\usepackage[authoryear]{natbib}
\usepackage[unicode=true]
 {hyperref}

\makeatletter

\providecommand{\tabularnewline}{\\}

\numberwithin{equation}{section}
\numberwithin{figure}{section}
\theoremstyle{plain}
\newtheorem{assumption}{\protect\assumptionname}
\theoremstyle{plain}
\newtheorem{thm}{\protect\theoremname}[section]

\usepackage{hyperref}
\usepackage{natbib} 
\usepackage[T1]{fontenc}
\usepackage{bm}

\newcommand\independent{\protect\mathpalette{\protect\independenT}{\perp}}
\def\independenT#1#2{\mathrel{\rlap{$#1#2$}\mkern2mu{#1#2}}}

\@ifundefined{showcaptionsetup}{}{%
 \PassOptionsToPackage{caption=false}{subfig}}
\usepackage{subfig}
\makeatother

\providecommand{\assumptionname}{Assumption}
\providecommand{\theoremname}{Theorem}

\begin{document}

\title{Semiparametric Estimation of a CES Demand System with \\
Observed and Unobserved Product Characteristics\thanks{We thank Jean-Pierre Dub{\'e}, \'Aureo de Paula, Thomas Chaney,
Steven T. Berry, Chad Syverson, Frank Wolak, Peter Newberry, Gaurab
Aryal, Robert J. LaLonde, Devesh Raval, Jakub Kastl, Mar Reguant,
Ying Fan, Naoki Aizawa, Alexander MacKay, and seminar participants
at the University of Chicago, IIOC 2016, Econometric Society NASM
2016, and SED 2016 for valuable comments and suggestions. Kyeongbae
Kim provided excellent research assistance. Horta\c{c}su gratefully
acknowledges financial support from the National Science Foundation
(SES 1426823). All errors are our own.}}

\author{Ali Horta\c{c}su\thanks{Department of Economics, University of Chicago, and NBER. \texttt{hortacsu@uchicago.edu}}
\and Joonhwi Joo\thanks{Naveen Jindal School of Management, University of Texas at Dallas.
\texttt{joonhwi.joo@utdallas.edu}} }

\date{June 5, 2018}

\maketitle
\thispagestyle{empty}
\begin{abstract}
We develop a characteristics based demand estimation framework for
the Marshallian demand system obtained by solving a budget-constrained
constant elasticity of substitution (CES) utility maximization problem.
From our Marshallian CES demand system, we derive the same market
share equation of \citet*{Berry1994,Berry1995}'s characteristics
based logit demand system. Our CES demand estimation framework can
accommodate zero predicted and observed market shares by conceptually
separating the whether-to-buy decision and how-much-to-buy decision.
Furthermore, the estimator we suggest allows a tractable semiparametric
estimation strategy that is flexible regarding the distribution of
unobservable product characteristics. We apply our framework to scanner
data on cola sales, where we show estimated demand curves can be upward
sloping if zero market shares are not accommodated properly.

JEL classification: C51, D11, D12 
\end{abstract}
\newpage{}

\setcounter{page}{1}
\begin{center}
{\LARGE{}Semiparametric Estimation of a CES Demand System with }\\
{\LARGE{}Observed and Unobserved Product Characteristics}{\LARGE\par}
\par\end{center}

\vspace{35bp}

\begin{abstract}
We develop a characteristics based demand estimation framework for
the Marshallian demand system obtained by solving a budget-constrained
constant elasticity of substitution (CES) utility maximization problem.
From our Marshallian CES demand system, we derive the same market
share equation of \citet*{Berry1994,Berry1995}'s characteristics
based logit demand system. Our CES demand estimation framework can
accommodate zero predicted and observed market shares by conceptually
separating the whether-to-buy decision and how-much-to-buy decision.
Furthermore, the estimator we suggest allows a tractable semiparametric
estimation strategy that is flexible regarding the distribution of
unobservable product characteristics. We apply our framework to scanner
data on cola sales, where we show estimated demand curves can be upward
sloping if zero market shares are not accommodated properly.

JEL classification: C51, D11, D12 
\end{abstract}
\newpage{}

\section{Introduction}

Constant elasticity of substitution (CES) preferences, often called
Dixit-Stiglitz-Spence preferences, have been used extensively to analyze
markets with product differentiation since \citet{Spence1976,Dixit1977,Anderson1979,Krugman1980}.
In marketing, demand systems derived from CES preference or its variants
have been used extensively in combination with mostly individual-
or purchase-level data (see, e.g., \citealp{Kim2002,Allenby2004,Dube2004,Kim2007,Lee2013,Lee2014,Howell2016}
among others). However, when only aggregate market-level data are
available to a researcher, \citet*{Berry1994,Berry1995}'s demand
estimation framework has become the \emph{de facto} standard method,
which is based on a different microfoundation \textendash{} the discrete
choice random utility model in the product characteristic space. We
reconcile these approaches of differentiated products demand estimation
using aggregate market-level data, showing that \emph{the direct utility
approach} based on CES preferences can also be just as rich and flexible
as \citet*{Berry1994,Berry1995}'s demand estimation framework. We
thereby shed light on an important connection between \emph{the direct
utility approach} and \emph{the indirect utility approach} (Section
3.1 of \citealp{Chintagunta2011}), or, stated differently, between
\emph{the neoclassical model} and \emph{the pure discrete choice model}
(Section 3 of \citealp{Dube2018}), in consumer demand estimation.

In this paper, we provide a general CES demand estimation framework
that can accommodate zero observed and predicted market shares. To
accommodate the zero shares, we develop a two-stage model of discrete-continuous
choice based on a budget-constrained CES utility maximization problem.
Our CES demand estimation framework is attractive for the following
reasons. First, we show that the identical market share equation of
\citet{Berry1994,Berry1995} can be derived from the budget-constrained
CES utility maximization problem. It allows the identification results
and estimation strategy developed for \citet{Berry1994,Berry1995}
to be directly applied to the CES demand system when zero market shares
are not present in the data. Second, when zero market shares are present
in the data, we explicitly introduce the exclusion restriction on
whether-to-buy decision of consumers, providing a conceptually clean
identification argument. Third, we employ a tractable semiparametric
estimation strategy that is flexible regarding the distribution of
unobservable product characteristics.

The current \emph{de facto} standard framework for differentiated
products demand estimation using aggregate market data was developed
by \citet{Berry1994,Berry1995}, which made breakthroughs in the demand
estimation literature in several aspects.\footnote{The breakthroughs include explicitly recognizing the correlation of
unobservable characteristics with the prices, market share inversion,
and simulation methods to estimate the random coefficients.} One of the breakthroughs was to (re)introduce the characteristic
space approach, which dates back to \citet{Lancaster1966}, in demand
estimation. The characteristic space approach can be very useful in
predicting the demand for a new product and evaluating its effects
on the market (\citealp{Petrin2002}). In \citet{Berry1994,Berry1995}'s
characteristics based demand estimation framework, a product is defined
as a bundle of observed and unobserved product characteristics. A
consumer can choose up to one product that yields the highest utility
among her finite choice set, or can decide to buy nothing. A consumer's
(dis)utility of consuming a product consists of the utility from price,
observed product characteristics, unobserved product characteristics,
and idiosyncratic utility shock. The individual choice probability
equation is derived from the distributional properties of the idiosyncratic
utility shock, which is assumed to follow the Type-I extreme value
distribution. Individual choice probabilities are taken as equal to
the predicted quantity shares of the individual demand, the aggregation
of which is taken as the predicted quantity market shares. We refer
to demand models based on these microfoundations as logit demand models,
which provide a tractable method of estimating differentiated product
demand systems by reducing the dimension of the parameters to be estimated. 

Our first contribution to the literature is provision of a concrete
link between the CES demand system and \citet{Berry1994,Berry1995}'s
homogeneous/random coefficients logit demand system. We do so by deriving
the identical market share equation of \citet{Berry1994,Berry1995}'s
characteristics based logit demand system from the CES demand system.
A nonnegative function in CES preferences that we refer to as the
\emph{quality kernel,} which is often referred to as the ``taste
parameter'' in the literature, plays a key role in directly incorporating
observed and unobserved product characteristics into the CES demand
system. Incorporating the taste parameter in CES preferences dates
back to at least \citet{Spence1976,Anderson1979}. To name just a
few, \citet{Kim2002,Dube2004} incorporated the idiosyncratic preference
shocks on the taste parameter. \citet{Einav2014} incorporated a sales
tax indicator and distance from the seller, which are the seller-consumer
specific characteristics, into the taste parameter of the CES preferences.
However, to the best of our knowledge, none of the literature models
the taste parameter of the CES preferences directly as a mapping from
the observed and unobserved product characteristics to develop a general
empirical framework for demand estimation. Adding the quality kernel
allows us to derive the identical, predicted quantity individual/market
share equation of \citet{Berry1994,Berry1995} from the resulting
Marshallian CES demand system. Early studies by \citet{Anderson1987,Anderson1992}
point out \emph{similarities} between the CES and logit demand systems
without product characteristics. Our market share equation equivalence
result is an extension of \citet{Anderson1987,Anderson1992}, in the
context of \citet{Berry1994,Berry1995}'s characteristics based demand
estimation framework. 

Our second contribution is the development of a direct method that
accommodates zero predicted and observed market shares. Accommodating
zero market shares in demand estimation has been a major difficulty
in the literature for decades, dating back to at least \citet{Deaton1980}.
\citet{Berry1994,Berry1995}'s logit demand model is not an exception.
Discrete choice frameworks with additive idiosyncratic errors with
unrestricted support inherently do not allow for zero individual choice
probabilities. Individual choice probabilities are treated as predicted
individual quantity shares, aggregated over homogeneous or heterogeneous
individuals and equated with observed market shares for identification
and estimation of model parameters. In logit demand models, additive
idiosyncratic shocks are distributed as an i.i.d. Type-I extreme value.
In such a case, the numerator of the individual choice probability
is the exponential of the alternative utility's deterministic part.
Provided that an alternative yields any utility higher than negative
infinity, the alternative must have a strictly positive predicted
market share. However, zero observed market shares are often observed
in data. Thus, in practice, researchers simply drop samples with zero
observed market shares or add a small, arbitrary number to zero observed
market shares. These \emph{ad hoc} measures cause biases in estimates.
We argue that selection in a consumer's consideration set must be
taken into account for identification and estimation of model parameters.
The consideration set selection drives the conditional expectation
of unobserved product characteristics that are conditioned on instruments
being non-zero and likely positive. The usual generalized method of
moments estimation yields price coefficient estimates that are biased
upward when this consideration set selection process is ignored.

To accommodate the zero predicted and observed market shares, we provide
a microfoundation for the selection-correction estimation equation
\emph{\textit{\`a la}} \citet{Heckman1979}, by embedding both extensive
and intensive margins on the quality kernel. \citet{Dubin1984,Hanemann1984,Chiang1991,Chintagunta1993,Kim2002,Nair2005}
introduced and developed modeling both margins in a utility maximization
problem to model the demand for a variety products. We extend the
idea to the Marshallian CES demand system and model a consumer's choice
as a two-stage decision process. In addition, we explicitly introduce
the exclusion restriction to the consideration set stage, which provides
a clean identification argument that separates the whether-to-buy
stage and how-much-to-buy stage.

Although we employ the direct utility approach, how we model the demand
for variety is different from most existing direct utility maximization
models in marketing. We model the demand for variety as a two-stage
decision, whether-to-buy and how-much-to-buy, in contrast to the one-step
decision on which the marketing literature has focused (see \citealp{Chintagunta2011,Dube2018}
for surveys). We interpret the first-stage whether-to-buy decision
as the consideration set formulation, which has a long tradition in
the marketing literature (see, e.g., \citet{Roberts1991,Ben-Akiva1995,Jedidi1996,Mehta2003,Gilbride2004}
among many others). It follows naturally that an item that never sold
in a market was not in the consideration set of consumers, which motivates
our use of the exclusion restriction and the selection-correction
estimation equation that accounts for the zero market shares. We employ
the \citet{Klein1993} estimator for the whether-to-buy stage, demonstrating
how the distribution-free efficient semiparametric estimator for the
binary response model can be easily applied to the demand estimation
problem with a multitude of zero predicted and observed market shares.
Furthermore, in our empirical example, we provide evidence that the
distribution of the unobservable product characteristics is far from
Gaussian. As for the contexts in which the single choice assumption
is more plausible, we also provide the microfoundation for the same
selection-correction estimation equation in the context of the logit
demand model.

In our empirical example in which our proposed estimation framework
is applied to scanner data that recorded the items on the shelves
that never sold during a given week, we demonstrate that dropping
zero market shares or imputing them with small positive numbers can
cause serious biases in the price-coefficient estimates. In particular,
if zero market shares are simply dropped, the price coefficient estimates
will be biased upward, even resulting in upward-sloping demand curves.
Our results have important implications for estimating demand elasticities
using scanner data: items that were on the shelves but not sold at
all must be considered and accommodated properly during estimation
of demand functions. Such information, however, is not included in
the majority of scanner datasets. We suggest that collecting such
information during the data collection procedure would be beneficial
for correctly estimating demand elasticities.

The remainder of the article is organized as follows. Section \ref{sec:Related-Literature}
briefly summarizes the related literature. Section \ref{sec:Microfoundation:-CES-Demand}
introduces our suggested CES demand system, and Section \ref{subsec:The-Exponential-Quality}
derives the market-share equation equivalence result. Section \ref{sec:Semiparametric-Estimation}
outlines the distribution-free semiparametric estimation framework
for our CES demand system, and Section \ref{sec:Derivation-of-Selection-Correcti}
provides a discussion relating our CES demand system to the pure discrete
choice logit demand system. Section \ref{sec:Monte-Carlo-Simulations}
provides Monte Carlo results, and Section \ref{sec:Empirical-Example}
presents an empirical example in which our framework is applied to
scanner data with a multitude of zero shares. Section \ref{sec:Conclusion}
concludes.

\section{\label{sec:Related-Literature}Related Literature}

This paper relates to the long tradition of consumer demand modeling
in the marketing literature, hedonic demand models in the economics
literature, and the economics literature accommodating zero observed
shares in multiplicative models.

In marketing, the direct utility approach was developed to accommodate
the purchase of multiple categories/brands. The most popular direct
utility specification would be translog preferences, which date back
to \citet{Christensen1975}. The richness of translog preferences
comes from the second-order term that allows for the possibility of
complements. In practice, however, second-order terms of translog
preferences are often omitted for the sake of tractability, in which
case the preferences can be nested as a variant of CES preference
\citep{Bhat2008}. \citet{Kim2002,Dube2004,Kim2007} used CES preference
or its variants in modeling the demand for a variety of consumers.
\citet{Allenby2004,Howell2016} used CES preference in the context
of nonlinear budget constraints associated with quantity discount
or price promotion, respectively, and \citet{Lee2013,Lee2014} used
it in the context of asymmetric complements and indivisibility of
demand.

Marketing literature has a long tradition of multiple discrete choice
and discrete-continuous models of demand that dates back to \citet{Hanemann1984}.
\citet{Chiang1991,Chintagunta1993,Mehta2007,Song2007} model quantity
choice and brand choice simultaneously, which entails the extensive
margins and intensive margins, respectively. Contrary to their focus
on the single brand choice, however, we focus on the zero observed
quantity shares and provide a tractable estimation method accommodating
the zero shares. The econometric model we derive is conceptually similar
to \citet{Gilbride2004,Nair2005}, where the screening and choice
components are modeled simultaneously. However, neither of them had
an exclusion restriction for discrete and quantity choice for identification,
and they do not account for zero market shares.

Logit models of consumer demand in marketing date back to at least
\citet{Guadagni1983}. When market-level data are available to the
researcher, the demand estimation framework developed by \citet{Berry1994,Berry1995}
has become the \emph{de facto} standard method to estimate the demand,
both in economics and marketing (see, e.g., \citealt{Besanko1998,Sudhir2001,Chintagunta2002,Chintagunta2005a,Draganska2006,Hitsch2006,Wilbur2008,Albuquerque2009,Goldfarb2009,Ghose2012}
among many others). The marketing literature suggests a few variants
to \citet{Berry1994,Berry1995} as well. \citet{Chintagunta2001}
developed an estimation method for when the idiosyncratic error term
is Gaussian, and \citet{Bruno2008} developed one for when the product
availability is varying. When individual-level data are available,
\citet{Villas-Boas1999,Chintagunta2005} used random coefficients
demand estimation with instrumental variables, to examine the effect
of correcting for endogeneity in brand choice.

Our approach to construct an empirical demand system without an idiosyncratic
preference shock can be viewed as a hedonic, or pure characteristics,
model of demand. Recent developments on hedonic demand estimation
frameworks were made by \citet{Bajari2005,Berry2007}, the former
of which relates more closely to our study. \citeauthor{Bajari2005}
investigate a general hedonic model of demand with product characteristics,
focusing on local identification and estimation of model parameters.
For global identification when a product space is continuous, they
specify Cobb-Douglas preferences. Our study extends their Cobb-Douglas
specification to the more flexible CES preferences specification that
can also accommodate zero predicted and observed market shares.%

Two papers in economics literature have tried to accommodate the zero
shares in the multiplicative models. \citet{Gandhi2013} rationalize
zero observed market shares differently, regarding such shares as
measurement errors of strictly positive predicted market shares, and
provide a partial identification result of model parameters. The difference
between their research and ours is that we rationalize zero predicted
and observed market shares, whereas they allow only observed market
shares to be zero. Nevertheless, their Monte Carlo simulations and
empirical applications suggest an implication similar to ours: when
samples with zero market shares are dropped, price coefficient estimates
are biased upward. In the international trade literature, \citet{Helpman2008}
developed another method that relates closely to ours in the context
of gravity models. They used a gravity model with endogenous censoring
of trade volumes, and their structural approach to handling zero trade
flows is similar to ours. However, their approach is fully parametric
in that they assume the Gaussian error term, whereas our approach
is semiparametric because we do not specify the distribution of unobservables
in our preferred specification.

\section{\label{sec:Microfoundation:-CES-Demand}CES Demand System with Observed
and Unobserved Product Characteristics}

\subsection{Specification of the CES Demand System}

We consider a differentiated product market denoted by subscript $t$,
composed of homogeneous consumers with a CES preference. We begin
by focusing on homogeneous consumers. The extension to product markets
comprised of heterogeneous consumers, with each consumer allowed to
have disparate utility parameters, is considered in Section \ref{subsec:Nesting-Homogenous-and}.
The utility from a product category is: 
\begin{equation}
u\left(\left\{ q_{j,t},\mathbf{x}_{j,t},\xi_{j,t},\mathbf{w}_{j,t},\eta_{j,t}\right\} _{j\in\mathcal{J}_{t}}\right):=\left(\sum_{j\in\mathcal{J}_{t}}\left\{ \chi\left(\mathbf{x}_{j,t},\xi_{j,t},\mathbf{w}_{j,t},\eta_{j,t}\right)\right\} ^{\frac{1}{\sigma}}q_{j,t}^{\frac{\sigma-1}{\sigma}}\right)^{\frac{\sigma}{\sigma-1}}.\label{eq:preference}
\end{equation}
Set $\mathcal{J}_{t}$ is a set of alternatives in the category, which
might include the numeraire that represents the outside option. $q_{j,t}$
is the quantity of product $j$ consumed in market $t$. $\chi\left(\mathbf{x}_{j,t},\xi_{j,t},\mathbf{w}_{j,t},\eta_{j,t}\right)$,
defined by the quality kernel, is a non-negative function of observed
and unobserved product characteristics. $\mathbf{x}_{j,t}$ and $\mathbf{w}_{j,t}$
are vectors of product $j$'s characteristics in market $t$, which
are observable to the econometrician. $\xi_{j,t}$ and $\eta_{j,t}$
are scalars that represent utility from product $j$'s characteristics
that are unobservable to the econometrician. $\mathbf{w}_{j,t}$ and
$\eta_{j,t}$ are extensive margin shifters that a consumer considers
whether to buy the product. $\mathbf{x}_{j,t}$ and $\xi_{j,t}$ are
intensive margin shifters that determine the level of utility when
a consumer buys a product. $\mathbf{w}_{j,t}$ and $\mathbf{x}_{j,t}$
might have common components, but we can require exclusion restriction
on $\mathbf{w}_{j,t}$ for semiparametric identification when the
extensive margin matters. In such a case, $\mathbf{w}_{j,t}$ must
contain at least one component that is not in $\mathbf{x}_{j,t}$.
We explain identification conditions further in Section \ref{sec:Semiparametric-Estimation}.
The observed extensive margin shifter, $\mathbf{w}_{j,t}$, might
contain the price $p_{j,t}$ or a nonlinear function of $p_{j,t}$.

The quality kernel $\chi\left(\mathbf{x}_{j,t},\xi_{j,t},\mathbf{w}_{j,t},\eta_{j,t}\right),$
introduced in equation (\ref{eq:preference}), is critical to our
framework. Researchers conventionally employ taste parameters or utility
weights in places we put the quality kernel. The quality kernel, taste
parameters, and utility weights can be commonly interpreted as multipliers
on the (marginal) utility of consuming a product. The quality kernel
is a straightforward extension of such conventions that allows us
to incorporate observed and unobserved product characteristics directly
into a consumer's utility. The quality kernel also allows the possibility
of explicitly separating intensive and extensive margins. This feature
accommodates zero predicted and observed market shares in model parameters.

The representative consumer's budget-constrained utility maximization
problem, the solution of which is the Marshallian demand system, is:
\begin{equation}
\max_{\left\{ q_{j,t}\right\} _{j\in\mathcal{J}_{t}}}\left(\sum_{j\in\mathcal{J}_{t}}\left\{ \chi\left(\mathbf{x}_{j,t},\xi_{j,t},\mathbf{w}_{j,t},\eta_{j,t}\right)\right\} ^{\frac{1}{\sigma}}q_{j,t}^{\frac{\sigma-1}{\sigma}}\right)^{\frac{\sigma}{\sigma-1}}\qquad s.t.\qquad\sum_{j\in\mathcal{J}_{t}}p_{j,t}q_{j,t}=y_{t}.\label{eq:zxkdjfas}
\end{equation}
The Marshallian demand system is: 
\begin{equation}
q_{j,t}=y_{t}\left\{ \frac{\chi\left(\mathbf{x}_{j,t},\xi_{j,t},\mathbf{w}_{j,t},\eta_{j,t}\right)p_{j,t}^{-\sigma}}{\sum_{k\in\mathcal{J}_{t}}\chi\left(\mathbf{x}_{k,t},\xi_{k,t},\mathbf{w}_{k,t},\eta_{k,t}\right)p_{k,t}^{1-\sigma}}\right\} \qquad\forall j\in\mathcal{J}_{t},\label{eq:optimalquantity}
\end{equation}
which leads to the predicted quantity market shares expression: 
\begin{eqnarray}
\pi_{j,t} & \equiv & \frac{q_{j,t}}{\sum_{k\in\mathcal{J}_{t}}q_{k,t}}\nonumber \\
 & = & \frac{\chi\left(\mathbf{x}_{j,t},\xi_{j,t},\mathbf{w}_{j,t},\eta_{j,t}\right)p_{j,t}^{-\sigma}}{\sum_{k\in\mathcal{J}_{t}}\chi\left(\mathbf{x}_{k,t},\xi_{k,t},\mathbf{w}_{k,t},\eta_{k,t}\right)p_{k,t}^{-\sigma}}.\label{eq:demandsystem}
\end{eqnarray}
Equation (\ref{eq:demandsystem}) is what we call the CES demand system
with observed and unobserved product characteristics. The demand system
(\ref{eq:demandsystem}), which is in the form of the predicted quantity
market shares, is our primary interest because the same predicted
quantity market share expression from \citet{Berry1994,Berry1995}
can be derived by imposing a further structure on the quality kernel,
$\chi\left(\cdot\right)$. (\ref{eq:demandsystem}), a system of predicted
quantity market shares, imposes only $\#\left(\mathcal{J}_{t}\right)-1$
constraints on the Marshallian demand system, $\mathbf{q}_{t}$, in
(\ref{eq:optimalquantity}). Only when combined with the budget constraint
$\sum_{j\in\mathcal{J}_{t}}p_{j,t}q_{j,t}=y_{t}$ can the Marshallian
demand quantities, $\mathbf{q}_{t}$, be uniquely pinned down for
a given price vector, $\left(\mathbf{p}_{t},y_{t}\right)\in\mathbb{R}^{\#\left(\mathcal{J}_{t}\right)+1}$.%

For the invertibility of the demand system, (\ref{eq:demandsystem}),
we consider the subset $\mathcal{J}_{t}^{+}\left(\subseteq\mathcal{J}_{t}\right)$,
such that $\pi_{j,t}>0$ for all $j\in\mathcal{J}_{t}^{+}$. The demand
system specified by $\left\{ \pi_{j,t}\right\} _{j\in\mathcal{J}_{t}^{+}}$
satisfies the connected substitutes conditions from \citet{Berry2013},
and is thus invertible. Invertibility of the demand system implies
that $\sigma$, the elasticity of substitution, is identified. If
we impose suitable structures on $\chi\left(\cdot\right)$, such as
monotonicity with index restriction, the structural parameters of
$\chi\left(\cdot\right)$ are also identified. We investigate the
specific functional forms of $\chi\left(\cdot\right)$ in Section
\ref{subsec:The-Exponential-Quality}.

\subsection{Properties of the CES Demand System and Comparison with the Logit
Demand System}

We now explain the properties of the CES demand system (\ref{eq:demandsystem}).
We begin with the Marshallian and Hicksian own and cross price elasticities
of the demand system. Let $b_{j,t}$ and $\pi_{j,t}$ be the budget
and quantity share of product $j$ in market $t$, respectively.\footnote{We use the term budget share and expenditure share exchangeably.}
Denote $\varepsilon_{jc,t}^{M}$ and $\varepsilon_{jc,t}^{H}$ by
the Marshallian and Hicksian cross price elasticities between alternatives
$j$ and $c$, respectively. If $\mathbf{w}_{j,t}$ does not include
the prices or function of the prices as its component, we have the
following simple closed-form formulas for the Marshallian and the
Hicksian own and cross price elasticities: 
\begin{eqnarray}
\varepsilon_{jj,t}^{M} & = & -\sigma+\left(\sigma-1\right)b_{j,t}\nonumber \\
\varepsilon_{jc,t}^{M} & = & \left(\sigma-1\right)b_{c,t}\nonumber \\
\varepsilon_{jj,t}^{H} & = & -\sigma\left(1-b_{j,t}\right)\nonumber \\
\varepsilon_{jc,t}^{H} & = & \sigma b_{c,t},\label{eq:priceelasticities}
\end{eqnarray}
and the income elasticity is 1.\footnote{In calculating the elasticities in practice, observed budget shares
can be used in place of $b_{j,t}$ and $b_{c,t}$.}\footnote{If $\mathbf{w}_{j,t}$ includes the prices or a function of the prices
so that the extensive margin is affected by the price changes, then
the simple closed-form expressions for the own and cross elasticities
cannot be derived. In practice, the corresponding price elasticities
can be calculated using simulations.} These elasticities can be easily calculated given that $\sigma$
is identified. From these elasticity expressions, it can be immediately
noticed that a version of the independence of irrelevant alternatives
(IIA) property holds; the substitution pattern depends solely on the
budget shares of corresponding products. The price elasticities of
the CES demand system should not be derived based on the quantity
market shares as in the logit demand models.\footnote{ 
\begin{eqnarray*}
\frac{\partial\ln\pi_{j,t}}{\partial\ln p_{j,t}} & = & \frac{\partial\left(\ln q_{j,t}-\ln\left(\sum_{k\in\mathcal{J}_{t}}q_{k,t}\right)\right)}{\partial\ln p_{j,t}}\\
 & = & \frac{\partial\ln q_{j,t}}{\partial\ln p_{j,t}}-\frac{\partial\ln\left(\sum_{k\in\mathcal{J}_{t}}q_{k,t}\right)}{\partial\ln p_{j,t}}\\
 & = & \varepsilon_{jj,t}^{M}-\frac{\partial\ln\left(\sum_{k\in\mathcal{J}_{t}}q_{k,t}\right)}{\partial\ln p_{j,t}}\\
 & \neq & \frac{\partial\ln q_{j,t}}{\partial\ln p_{j,t}}
\end{eqnarray*}
The term $\frac{\partial\ln\pi_{j,t}}{\partial\ln p_{j,t}}$ is the
Marshallian price elasticity only when $\sum_{k\in\mathcal{J}_{t}}q_{k,t}$
is constant, which is the case for the logit demand models. See Appendix
\ref{sec:Derivation-of-the} for the details.}  Price elasticities in the logit demand models when the mean utility
is log-linear in prices are given by: 
\begin{eqnarray}
\varepsilon_{jj,t}^{L} & = & -\alpha\left(1-\pi_{j,t}\right)\nonumber \\
\varepsilon_{jc,t}^{L} & = & \alpha\pi_{c,t}.\label{eq:logitelasticity}
\end{eqnarray}
\footnote{When the mean utility is linear in prices, the elasticity expression
becomes: 
\begin{eqnarray*}
\varepsilon_{jj,t}^{L} & = & -\alpha p_{j,t}\left(1-\pi_{j,t}\right)\\
\varepsilon_{jc,t}^{L} & = & \alpha p_{c,t}\pi_{c,t}.
\end{eqnarray*}
See Section \ref{subsec:Nesting-Homogenous-and} for a further discussion
on the functional form of the mean utilities in the logit demand model.}\footnote{In calculating the elasticities in practice, observed quantity shares
can be used in place of $\pi_{j,t}$ and $\pi_{c,t}$.} The expressions (\ref{eq:logitelasticity}) parallel the Hicksian
price elasticities of the CES demand system. The only difference to
the Hicksian price elasticities (\ref{eq:priceelasticities}), derived
from the CES demand system, is that the multiplied terms of the log-price
coefficient, $\alpha$, are comprised of quantity market shares, not
budget shares. 

Because we derive the demand system from the budget-constrained CES
utility maximization problem, the duality between the Marshallian
and Hicksian demand functions holds. The Slutsky equation follows,
and thus we can decompose the substitution and income effect more
naturally. The Slutsky equation in the elasticity form is: 
\[
\varepsilon_{jc,t}^{M}=\varepsilon_{jc,t}^{H}-\varepsilon_{j,t}^{I}b_{c,t}.
\]
Because $\varepsilon_{j,t}^{I}=1$ in the CES demand system, the income
effect depends solely on budget shares, which is a considerable limitation.
However, there are at least two advantages over the discrete choice
counterpart. First, although the numeraire can be included in the
consumer's consideration set, $\mathcal{J}_{t}$, it is unnecessary
in our CES demand system. In contrast, inclusion of the numeraire
in the consideration set is necessary in the logit demand system to
induce an income effect, especially when the income is not a direct
argument of the discrete choice utility or it is canceled out.\footnote{For example, the case when the mean utility is linear in income net
of price.} In such a case, the price increase of an alternative leads to consumers
switching to only other alternatives in the consideration set. The
magnitude of the income effect is in a sense determined \emph{a priori}
by the researcher in logit demand models because the income effect
depends primarily on quantity market shares of the numeraire. The
size of the share of the numeraire is often arbitrarily assumed or
imposed by a researcher in practice. Second, the income effect depends
on budget shares, not quantity shares, in the Marshallian CES demand
system. In logit demand models, the income effect of a product with
a small budget share and a large quantity share is large, which is
even more unrealistic.

\section{\label{subsec:The-Exponential-Quality}The Exponential Quality Kernel}

So far, we have not restricted the quality kernel, $\chi\left(\mathbf{x}_{j,t},\xi_{j,t},\mathbf{w}_{j,t},\eta_{j,t}\right)$.
In principle, $\chi\left(\mathbf{x}_{j,t},\xi_{j,t},\mathbf{w}_{j,t},\eta_{j,t}\right)$
can be any non-negative function. Under this weak restriction, the
demand system specified by predicted market shares (\ref{eq:demandsystem})
can be identified locally, as investigated by \citet{Bajari2005}.
However, nonparametric estimation of a locally identified demand system
places a considerable burden on the data and computational power,
which is often impractical. Locally identified parameter values are
often uninformative regarding counterfactual analyses, and alternatively,
we can impose further structures on the consumer utility from product
characteristics. We focus on the exponential quality kernel with an
index restriction. This specific functional form deserves a special
attention for two reasons. First, by using this functional form, we
can derive the same individual choice probability equation of the
homogeneous and random coefficient logit models of demand from the
CES demand system developed in the previous section. Second, this
functional form simplifies the estimation problem substantially because
the estimation equation reduces to a log-linear form. We use the exponential
quality kernel to propose a tractable, semiparametric estimation method
that accommodates zero predicted and observed market shares.

\subsection{\label{subsec:Nesting-Homogenous-and}Nesting the Homogeneous and
Random Coefficient Logit Models of Demand}

We show that the predicted quantity market share expressions of the
homogeneous and random coefficient logit models of demand can be derived
from (\ref{eq:demandsystem}) by choosing a functional form of the
quality kernel, $\chi\left(\mathbf{x}_{j,t},\xi_{j,t},\mathbf{w}_{j,t},\eta_{j,t}\right)$.
Suppose that $\mathbf{x}_{j,t}=\mathbf{w}_{j,t}$, $\xi_{j,t}=\eta_{j,t}$,
$\chi\left(\mathbf{x}_{j,t},\xi_{j,t}\right)>0$, $\pi_{j,t}>0$,
and let $\mathbf{x}_{j,t}$ be exogenous for all $j,t$. We do not
require an exclusion restriction in this setup because the predicted
quantity shares are positive for every alternative. Let $\mathcal{J}_{t}$
contain the numeraire, denoted by product $0$, and normalize $p_{0,t}=1$.\footnote{We emphasize that $\mathcal{J}_{t}$ might not contain a numeraire
for our CES demand system. In such a case, product $0$ can be considered
any alternative in $\mathcal{J}_{t}$, and all estimation equations
that follow should be adjusted in terms of differences between $j$
and $0$. } Taking the ratios of products $j$ and $0$, and taking the logarithm
of equation (\ref{eq:demandsystem}), yields: 
\begin{equation}
\ln\left(\frac{\pi_{j,t}}{\pi_{0,t}}\right)=-\sigma\ln\left(p_{j,t}\right)+\ln\chi\left(\mathbf{x}_{j,t},\xi_{j,t}\right)-\ln\chi\left(\mathbf{x}_{0,t},\xi_{0,t}\right).\label{eq:logit}
\end{equation}
We normalize $\mathbf{x}_{0,t}=\mathbf{0}$, $\xi_{0,t}=0$, and let
$\chi\left(\mathbf{x}_{j,t},\xi_{j,t}\right)=\exp\left(\mathbf{x}_{j,t}'\bm{\beta}+\xi_{j,t}\right)$.
(\ref{eq:logit}) then becomes: 
\begin{equation}
\ln\left(\frac{\pi_{j,t}}{\pi_{0,t}}\right)=-\sigma\ln\left(p_{j,t}\right)+\mathbf{x}_{j,t}'\bm{\beta}+\xi_{j,t}.\label{eq:logitestimation}
\end{equation}
(\ref{eq:logitestimation}) coincides with the estimation equation
of the homogeneous logit model of demand, except that in (\ref{eq:logitestimation}),
$\ln\left(p_{j,t}\right)$ is used in place of $p_{j,t}$, which is
a convention in the literature.\footnote{See Appendix \ref{sec:Derivation-of-the} for the derivation of homogeneous
and random coefficient logit models.} The log of price should be used in (\ref{eq:logitestimation}) because
it is inherited from the consumer's budget constraint. In contrast,
we observe that $\ln\left(p_{j,t}\right)$ can also be used in place
of $p_{j,t}$ in the utility specification of the logit demand system;
by substituting $\ln\left(p_{j,t}\right)$ with $p_{j,t}$ in the
linear utility specification in the logit demand model, the estimation
equation of the proposed CES demand system lines up exactly with that
of the homogeneous logit demand system. We take this substitution
with the log of prices as a simple scale adjustment in the linear
utility specification of the logit demand model. The predicted market
share equation of the random coefficient logit model of demand developed
by \citet{Berry1995} can be derived similarly. Let $i$ denote an
individual, and suppress the market subscript $t$ temporarily. For
the sake of notational simplicity, let $\phi_{j}:=\ln p_{j}$. We
specify the quasi-linear utility of the random coefficient logit model
of demand as: 
\[
u_{i,j}=-\alpha_{i}\phi_{j}+\mathbf{x}_{j}'\bm{\beta}_{i}+\xi_{j}+\epsilon_{i,j}.
\]
In contrast, the individual quantity share expressions of the CES
demand system (\ref{eq:demandsystem}) become: 
\begin{eqnarray}
\pi_{i,j} & = & \frac{\chi_{i}\left(\mathbf{x}_{j},\xi_{j}\right)\exp\left(-\sigma_{i}\phi_{j}\right)}{\sum_{k=0}^{J}\chi_{i}\left(\mathbf{x}_{k},\xi_{k}\right)\exp\left(-\sigma_{i}\phi_{k}\right)}\label{eq:djkasdfs}\\
 & = & \frac{\exp\left(-\sigma_{i}\phi_{j}+\mathbf{x}_{j}'\bm{\beta}_{i}+\xi_{j}\right)}{\sum_{k=0}^{J}\exp\left(-\sigma_{i}\phi_{k}+\mathbf{x}_{k}'\bm{\beta}_{i}+\xi_{k}\right)},\label{eq:asdjkfasd}
\end{eqnarray}
where the second equality follows by specifying $\chi_{i}\left(\mathbf{x}_{j},\xi_{j}\right)=\exp\left(\mathbf{x}_{j}'\bm{\beta}_{i}+\xi_{j}\right)$.
Note that (\ref{eq:asdjkfasd}) is nearly identical to the individual
choice probability equation obtained by \citet{Berry1995}.\footnote{The only structural difference is the correlation structure of the
individual heterogeneity; we must assume that $Cov\left(\sigma_{i},\bm{\beta}_{i}\right)=\mathbf{0}$.
As those cross-correlations are often assumed to be zero in practice
when estimating the random coefficient logit model of demand (see
\citet{Dube2012}), we do not consider the restriction a serious limitation. } The predicted market share equation is obtained by aggregating these
individual quantity shares over $i$.

Discussions in the current subsection provide the microfoundation
and justification for international trade and macroeconomics literature,
based on the CES demand system, to use differentiated products demand
estimation methods developed in empirical industrial organizational
literature since \citet{Berry1994,Berry1995,Nevo2001}. After model
parameters are estimated, price and income elasticities can be calculated
according to equation (\ref{eq:priceelasticities}), and the welfare
analyses can be conducted correspondingly. 

However, discrete choice differentiated product demand estimation
literature imposes a critical restriction, which is necessary when
inverting the individual quantity share, $\pi_{j,t}>0$, for all $j,t$.\footnote{For a detailed discussion on share inversion, see \citet{Berry2013}.}
The restriction is inevitable in logit demand models, which assume
additive idiosyncratic shocks on preferences distributed with unrestricted
support. The most important example in the literature is additive
i.i.d. Type-I extreme value distributed shocks. Individual choice
probabilities derived from the assumption must have exponential functions
in the numerators of choice probabilities. Zero quantity market shares
are often observed in data, which are equated with predicted market
shares for identification and estimation of model parameters. The
flexibility of the quality kernel, $\chi\left(\mathbf{x}_{j,t},\xi_{j,t},\mathbf{w}_{j,t},\eta_{j,t}\right)$,
in our model allows us to accommodate zero predicted market shares
by embedding a buy-or-not decision of the consumer, which determines
extensive margins. We now illustrate how to accommodate zero predicted
and observed market shares directly.

\subsection{\label{subsec:Introducing-Quality-Kernel}Accommodating Zero Predicted
and Observed Market Shares: Separating Intensive and Extensive Margins}

We restrict attention to homogeneous consumers again, and let $\mathbf{x}_{j,t}\neq\mathbf{w}_{j,t}$,
$\eta_{j,t}\neq\xi_{j,t}$. We let $\mathcal{J}_{t}$ contain the
numeraire for convenience of illustration, and normalize $p_{0,t}=1$.
The predicted market shares equation of the proposed CES demand system
is: 
\begin{eqnarray}
\pi_{j,t} & = & \frac{\chi\left(\mathbf{x}_{j,t},\xi_{j,t},\mathbf{w}_{j,t},\eta_{j,t}\right)\exp\left(-\sigma\phi_{j,t}\right)}{\sum_{k\in\mathcal{J}_{t}}\chi\left(\mathbf{x}_{j,t},\xi_{j,t},\mathbf{w}_{j,t},\eta_{j,t}\right)\exp\left(-\sigma\phi_{k,t}\right)}.\label{eq:individualprob}
\end{eqnarray}
The expression (\ref{eq:individualprob}) allows zero predicted market
shares of product $j$ by letting $\chi\left(\mathbf{x}_{j,t},\xi_{j,t},\mathbf{w}_{j,t},\eta_{j,t}\right)=0$
for some subset of the product characteristic space where $\left(\mathbf{w}_{j,t},\eta_{j,t}\right)$
lives on. By taking the ratio $\pi_{j,t}/\pi_{0,t}$, we obtain a
reduced form of the demand system (\ref{eq:individualprob}) as: 
\begin{equation}
\frac{\pi_{j,t}}{\pi_{0,t}}=p_{j,t}^{-\sigma}\frac{\chi\left(\mathbf{x}_{j,t},\xi_{j,t},\mathbf{w}_{j,t},\eta_{j,t}\right)}{\chi\left(\mathbf{x}_{0,t},\xi_{0,t},\mathbf{w}_{0,t},\eta_{0,t}\right)}.\label{eq:demandsystem2}
\end{equation}
If $\mathcal{J}_{t}$ does not include the numeraire, any product
with a strictly positive market share can be considered a reference
product, denoted by product $0$. All arguments in the current and
subsequent sections remain valid provided that statistical independence
of the observable and unobservable product characteristics across
products can be assumed. This assumption implies that product characteristics
are uncorrelated across products, which is consistent with many extant
demand estimation frameworks, including \citet{Berry1994,Berry1995}.
For tractability during identification and estimation, we consider
the following functional form with an index restriction: 
\begin{eqnarray}
\chi\left(\mathbf{x}_{j,t},\xi_{j,t},\mathbf{w}_{j,t},\eta_{j,t}\right) & = & \mathbf{1}\left(\left\{ \gamma+\mathbf{w}_{j,t}'\bm{\delta}+\eta_{j,t}>0\right\} \right)\exp\left(\alpha+\mathbf{x}_{j,t}'\bm{\beta}+\xi_{j,t}\right),\label{eq:chi}
\end{eqnarray}
where $\mathbf{1}\left(\cdot\right)$ is an indicator function. Employing
this quality kernel is equivalent to assuming a certain structure
on the consumer's choice. A consumer initially considers the utility
from product characteristics represented by $\mathbf{w}_{j,t}'\bm{\delta}+\eta_{j,t}$.
If the utility exceeds the threshold $-\gamma$, the consumer decides
to buy the product. Then $\left(\phi_{j,t},\mathbf{x}_{j,t},\xi_{j,t}\right)$
is considered, which affects the amount of consumption $q_{j,t}$.
In contrast, if the utility does not exceed the threshold $-\gamma$,
the consumer decides not to buy the product, and thus, $q_{j,t}=\pi_{j,t}=0$.
We emphasize that $\mathbf{w}_{j,t}$ can contain the raw price, $p_{j,t}$,
or other endogenous variables provided that the corresponding instruments
are available to the researcher.

\section{\label{sec:Semiparametric-Estimation}A Semiparametric Estimation
Framework with Exponential Quality Kernel and Zero Market Shares}

We provide a semiparametric estimation framework for the CES demand
system with exponential quality kernel that accommodates zero predicted
and observed market shares. The estimation method we provide includes
two stages. During the first stage, parameters that determine extensive
margins are estimated using the efficient semiparametric estimator
developed by \citet{Klein1993}, and during the second, parameters
that determine intensive margins are estimated, correcting for price
endogeneity and selectivity bias caused by a consumer's consideration
set selection. The second-stage estimator that we use was developed
by \citet{Ahn1993,Powell2001}. When zero market shares are not observed
in the data, one can proceed with existing demand estimation frameworks
developed by \citet{Berry1994,Berry1995} to estimate model parameters.
The first-stage estimation framework illustrated in this section allows
only exogenous covariates for the observed extensive margin shifter,
$\mathbf{w}_{j,t}$. We chose this framework because of the availability
of data and efficiency.\footnote{Although our data include information on product availability to consumers,
even when sales in a corresponding week/store pair were zero, they
do not include prices in corresponding weeks/stores without sales.
Thus, we could not contain the endogenous variable $p_{j,t}$ in $\mathbf{w}_{j,t}$
during first-stage estimation. Characteristics of data used in our
empirical application are discussed in Section \ref{sec:Empirical-Example}. } If a researcher wants to include endogenous variables such as prices
in the extensive margin shifters, $\mathbf{w}_{j,t}$, the researcher
can proceed with the method developed by \citet{Blundell2003,Blundell2004}
or \citet{Rothe2009} during the first stage. They provide semiparametric
estimation frameworks for binary choice models with endogenous covariates.

We assume the existence of instruments for prices, such that $E\left[\xi_{j,t}|\mathbf{z}_{j,t}\right]=0$,
where $\mathbf{z}_{j,t}$ might include $\mathbf{x}_{j,t}$. Let $d_{j,t}=\mathbf{1}\left(\left\{ \gamma+\mathbf{w}_{j,t}'\bm{\delta}+\eta_{j,t}>0\right\} \right)$.
It is well documented in the literature that $E\left[\xi_{j,t}|\phi_{j,t},\mathbf{x}_{j,t}\right]\neq0$,
and it is highly likely to be positive. Consequently, when prices
are not instrumented, upward-sloping demand curves are often estimated.
The same intuition applies when a consumer's consideration set selection
is ignored and samples with zero observed market shares are simply
dropped during estimation. Even after instrumenting for prices, $E\left[\xi_{j,t}|\mathbf{z}_{j,t}\right]=0$
does not imply that $E\left[\xi_{j,t}|\mathbf{z}_{j,t},d_{j,t}=1\right]$
is zero. $E\left[\xi_{j,t}|\mathbf{z}_{j,t},d_{j,t}=1\right]$ is
likely to be positive because consumers select products with high
$\eta_{j,t}$ during the first-stage consideration set decision, and
$\eta_{j,t}$ is likely to be positively correlated with $\xi_{j,t}$.
Thus, dropping samples with zero observed market shares during estimation
biases price coefficients upward, which can even yield positive price
coefficients. Imputing zero observed market shares with some small
positive numbers during estimation can cause an even more serious
problem in that the direction of the bias is unpredictable. 

We normalize $\phi_{0,t}\equiv\ln p_{0,t}=0$, $\xi_{0,t}=\eta_{0,t}=0$,
$\mathbf{w}_{0,t}=\mathbf{0}$, and $\mathbf{x}_{0,t}=\mathbf{0}$.
Under the choice of $\chi\left(\cdot\right)$ specified in (\ref{eq:chi}),
(\ref{eq:demandsystem2}) simplifies to: 
\begin{equation}
\frac{\pi_{j,t}}{\pi_{0,t}}=\mathbf{1}\left(\left\{ \gamma+\mathbf{w}_{j,t}'\bm{\delta}+\eta_{j,t}>0\right\} \right)\exp\left(-\sigma\phi_{j,t}+\mathbf{x}_{j,t}'\bm{\beta}+\xi_{j,t}\right),\label{eq:zxdfhaksdf}
\end{equation}
which is the econometric model that we identify and estimate in this
section. A consumer buys product $j$ if $\gamma+\mathbf{w}_{j,t}'\bm{\delta}+\eta_{j,t}>0$.
For the sample with $d_{j,t}=1$, demand system (\ref{eq:zxdfhaksdf})
further reduces to: 
\[
\ln\left(\frac{\pi_{j,t}}{\pi_{0,t}}\right)=-\sigma\phi_{j,t}+\mathbf{x}_{j,t}'\bm{\beta}+\xi_{j,t}.
\]
However, the conditional expectation $E\left[\xi_{j,t}|\mathbf{z}_{j,t},\mathbf{w}_{j,t},d_{j,t}=1\right]$
is not zero anymore, which leads to the sample selection problem.
Several methods to estimate parameters of the sample selection models
have been proposed in the literature under different assumptions.\footnote{For example, \citet{Powell1984,Powell1986,Blundell2007} propose a
least absolute deviation type estimator under the conditional quantile
restriction, and \citet{Honore1997} propose symmetric trimming under
the symmetricity assumption of error terms. } We follow \citet{Heckman1979}, who imposes a conditional mean restriction.
By taking the conditional expectation, we have: 
\begin{equation}
E\left[\ln\left(\frac{\pi_{j,t}}{\pi_{0,t}}\right)|\mathbf{z}_{j,t},\mathbf{w}_{j,t},d_{j,t}=1\right]=-\sigma\phi_{j,t}+\mathbf{x}_{j,t}'\bm{\beta}+E\left[\xi_{j,t}|\mathbf{z}_{j,t},\mathbf{w}_{j,t},d_{j,t}=1\right].\label{eq:loglineardemand}
\end{equation}
\citet{Ahn1993,Powell2001,Newey2009} propose two-stage $\sqrt{N}$-consistent
estimators for the model parameters of (\ref{eq:loglineardemand}).
We use the pairwise differenced weighted least squares estimator from
\citet{Ahn1993,Powell2001}, which corrects for the endogeneity of
$\phi_{j,t}$ using instruments during the second stage. During the
first stage, $\bm{\delta}$ should be estimated. A few estimators
are available for this semiparametric binary choice model, among which
we use the method from \citet{Klein1993}, which achieves asymptotic
efficiency. During the second stage, parameters $\left(\sigma,\bm{\beta}\right)$
from the following linear equation are estimated: 
\begin{equation}
E\left[\ln\left(\frac{\pi_{j,t}}{\pi_{0,t}}\right)|\mathbf{z}_{j,t},\mathbf{w}_{j,t},d_{j,t}=1\right]=-\sigma\phi_{j,t}+\mathbf{x}_{j,t}'\bm{\beta}+\lambda\left(1-G_{\eta}\left(-\mathbf{w}_{j,t}'\bm{\delta}\right)\right),\label{eq:loglinear_estimation_eq}
\end{equation}
where $\lambda\left(\cdot\right)$ is an unknown smooth function.
For semiparametric identification of $\left(\sigma,\bm{\beta}\right)$,
term $\lambda\left(1-G_{\eta}\left(-\mathbf{w}_{j,t}'\bm{\delta}\right)\right)$
must not be a linear combination of $\left(\phi_{j,t},\mathbf{x}_{j,t}\right)$;
some component of $\mathbf{w}_{j,t}$ must be excluded from $\left(\phi_{j,t},\mathbf{x}_{j,t}\right)$.
We impose the following assumptions on the data-generating process
for the $\sqrt{N}$-consistency and asymptotic normality of our proposed
estimator.
\begin{assumption}
\label{assu:The-product-characteristic}The vector of observed product
characteristics $\mathbf{w}_{j,t}$ is exogenous.
\end{assumption}

\begin{assumption}
\label{assu:exclusion}$\mathbf{w}_{j,t}$ contains at least one component
that is not included in $\mathbf{x}_{j,t}$.
\end{assumption}

\begin{assumption}
\label{assu:-is-independent}$\eta_{j,t}$ is independent of $\mathbf{w}_{j,t}$
with $E\left[\eta_{j,t}|\mathbf{w}_{j,t}\right]=0$, $\eta_{j,t}$
is i.i.d. over $j$ and over $t$, and the conditional distribution
of $\eta_{j,t}$ given $\mathbf{w}_{j,t}$ has the full support over
$\mathbb{R}$ with bounded first derivatives.
\end{assumption}

\begin{assumption}
\label{assu:-and-}$\mathbf{w}_{j,t}$ and $\mathbf{x}_{j,t}$ contain
at least one continuous variable. Furthermore, the conditional distribution
of the continuous variable conditioned on $d_{j,t}$ and other exogenous
variables is sufficiently smooth.
\end{assumption}

\begin{assumption}
\label{assu:Denote--by}Denote $\mathbf{r}_{j}:=\left(\phi_{j},\mathbf{x}_{j}\right)'$.
Denote $g_{\left(\cdot\right)}\left(w\right):=E\left[\cdot|\mathbf{w}_{j,t}'\bm{\delta}=w\right]$
and $f\left(w\right)$ be the density of $\mathbf{w}_{j,t}'\bm{\delta}$.
Then, $f\left(w\right)$, $g_{\mathbf{r}_{j,t}}\left(w\right)$, $g_{\mathbf{z}_{j,t}}\left(w\right)$,
$g_{\eta_{j,t}}\left(w\right)$, $g_{\mathbf{z}_{j,t}\mathbf{w}_{j,t}'}\left(w\right)$,
and $g_{\mathbf{z}_{j,t}\mathbf{r}_{j,t}'}\left(w\right)$ are sufficiently
smooth on their supports.
\end{assumption}

\begin{assumption}
\label{assu:There-exists-a}There exists a set of instruments $\mathbf{z}_{j,t}$
such that $\xi_{j,t}\independent\phi_{j,t}|\mathbf{z}_{j,t}$, $E\left[\xi_{j,t}|\mathbf{z}_{j,t}\right]=0$,
and $\dim\left(\mathbf{z}_{j,t}\right)\geq\dim\left(\phi_{j,t},\mathbf{x}_{j,t}\right)$.\footnote{$\mathbf{z}_{j,t}$ may contain the exogenous components of $\mathbf{x}_{j,t}$.}
\end{assumption}

\begin{assumption}
\label{assu:The-parameter-vector}The parameter vector $\left(\sigma,\alpha,\bm{\beta},\gamma,\bm{\delta}\right)$
lies in a compact parameter space, with the true parameter value lying
in the interior.
\end{assumption}
In Assumptions \ref{assu:The-product-characteristic} through \ref{assu:-is-independent},
we impose the independence of observed and unobserved product characteristics,
and homoskedasticity of unobservable product characteristics, $\eta_{j,t}$,
that relate to extensive margins. However, we do not assume that unobserved
product characteristics and prices are independent. We allow for endogeneity
in prices, which should be considered during identification and estimation;
prices can be a function of observed and unobserved product characteristics.
Assumption \ref{assu:exclusion} is the exclusion restriction, which
is required for identification during the second stage. Note that
a sufficient condition for Assumption \ref{assu:-is-independent}
to hold is that $\eta_{j,t}\independent\mathbf{w}_{j,t}$ and $\eta_{j,t}$
has a bounded and continuous density over the real line. Assumptions
\ref{assu:-and-} and \ref{assu:Denote--by} are smoothness conditions,
imposed for the suggested estimators to be well-behaved.\footnote{Roughly, they require the existence of the higher-order derivatives
for the respective conditional distribution and conditional expectation
functions. See (C.6) of \citet{Klein1993} and Assumption 5.7 of \citet{Powell2001}
for the exact conditions.} Assumption \ref{assu:There-exists-a} is the standard instrument
condition to correct for price endogeneity.\footnote{See, e.g., \citet{Nevo2001} for a discussion of suitable instruments
in practice.} Assumption \ref{assu:The-parameter-vector} is the usual compactness
assumption.%
{} 

We now describe first- and second-stage estimators. During the first
stage, we estimate $\bm{\delta}$ using the efficient semiparametric
estimator developed by \citet{Klein1993}. The estimator allows us
to estimate parameters of binary choice models without having to specify
the distribution of the unobservables. The insight is to replace the
likelihood with its uniformly consistent estimates, and run the pseudo-maximum
likelihood. The estimator is defined as: 
\begin{equation}
\hat{\bm{\delta}}:=\arg\max_{\bm{\delta}}\sum_{j,t}\left\{ \mathbf{1}\left(\pi_{j,t}>0\right)\ln\left(1-\hat{G}_{\eta}\left(-\mathbf{w}_{j,t}'\bm{\delta}\right)\right)+\mathbf{1}\left(\pi_{j,t}=0\right)\ln\left(\hat{G}_{\eta}\left(-\mathbf{w}_{j,t}'\bm{\delta}\right)\right)\right\} ,\label{eq:ks_obj}
\end{equation}
where 
\[
\hat{G}_{\eta}\left(-\mathbf{w}_{j,t}'\bm{\delta}\right)=\hat{\tau}_{j,t}\frac{\sum_{k\neq j,t}\kappa\left(\frac{1}{h_{n}}\left(\mathbf{w}_{k}-\mathbf{w}_{j,t}\right)'\bm{\delta}+\iota_{0}\left(\bm{\delta}\right)\right)\left(1-\mathbf{1}\left(\pi_{j,t}>0\right)\right)}{\sum_{k\neq j,t}\kappa\left(\frac{1}{h_{n}}\left(\mathbf{w}_{k}-\mathbf{w}_{j,t}\right)'\bm{\delta}+\iota\left(\bm{\delta}\right)\right)}.
\]
$\kappa\left(\cdot\right)$ is a fourth-order kernel, $h_{n}$ is
the bandwidth, and $\hat{\tau}_{j,t},\iota_{0}\left(\bm{\delta}\right),\iota\left(\bm{\delta}\right)$
are trimming sequences for small estimated densities.\footnote{We ignore these trimming sequences for technical and notational convenience
from now on. \citet{Klein1993} also note that the trimming does not
affect the estimates in practice.} During the second stage, we follow the method illustrated by \citet{Powell2001}.
With an abuse of notation by suppressing the market index $t$ and
letting $\mathbf{r}_{j}=\left(\phi_{j},\mathbf{x}_{j}\right)'$, the
estimator is defined by the following weighted instrumental variable
estimator: 
\begin{equation}
\left(-\hat{\sigma},\hat{\bm{\beta}}\right)=\left(\sum_{i=1}^{n-1}\sum_{j=i+1}^{n}\hat{\omega}_{i,j}\left(\mathbf{z}_{i}-\mathbf{z}_{j}\right)\left(\mathbf{r}_{i}-\mathbf{r}_{j}\right)'\right)^{-1}\left(\sum_{i=1}^{n-1}\sum_{j=i+1}^{n}\hat{\omega}_{i,j}\left(\mathbf{z}_{i}-\mathbf{z}_{j}\right)\left(\ln\left(\frac{\pi_{i}}{\pi_{0}}\right)-\ln\left(\frac{\pi_{j}}{\pi_{0}}\right)\right)\right),\label{eq:betahat}
\end{equation}
where $\hat{\omega}_{i,j}=\frac{1}{h_{n}}\kappa\left(\frac{1}{h_{n}}\left(\mathbf{w}_{i}-\mathbf{w}_{j}\right)'\hat{\bm{\delta}}\right)$.\footnote{When the number of instruments is larger than that of explanatory
variables, the projection matrix can be calculated beforehand to find
the $\mathbf{z}_{j}$ vector. Efficiency loss might occur, but the
estimator will be still $\sqrt{N}$-consistent and asymptotically
normal.}\footnote{The bandwidth sequence $h_{n}$ should be such that $h_{n}\rightarrow0$,
$nh_{n}^{6}\rightarrow\infty$, and $nh_{h}^{8}\rightarrow0$ as $n\rightarrow\infty$
in both the first and the second stage.} Intuition regarding the estimator suggests canceling out the bias
correction term, $\lambda\left(1-G_{\eta}\left(-\mathbf{w}_{j}'\bm{\delta}\right)\right)$;
if $\mathbf{w}_{i}$ is the same as $\mathbf{w}_{j}$, term $\lambda\left(1-G_{\eta}\left(-\mathbf{w}_{j}'\bm{\delta}\right)\right)$
in (\ref{eq:loglinear_estimation_eq}) cancels out when differences
are taken. Thus, more weights are placed on differenced terms that
are close. The estimator is $\sqrt{N}$-consistent and asymptotically
normal. For the closed-form covariance matrix formula and its consistent
estimator, see \citet{Powell2001}. The following theorem summarizes
discussions in this subsection thus far.
\begin{thm}
Under Assumptions \ref{assu:The-product-characteristic} through \ref{assu:The-parameter-vector},
$\left(-\hat{\sigma},\hat{\bm{\beta}}\right)$, defined in (\ref{eq:betahat}),
is $\sqrt{N}$-consistent and asymptotically normal.
\end{thm}
The semiparametric, log-linear estimation illustrated in this subsection
requires an exclusion restriction on $\mathbf{w}_{j,t}$ to identify
$\left(-\sigma,\bm{\beta}\right)$; $\mathbf{w}_{j,t}$ cannot be
a linear combination of $\left(\phi_{j,t},\mathbf{x}_{j,t}\right)$.
This exclusion restriction can be circumvented by adding an interaction
term or nonlinear transformation of a non-binary variable contained
in $\left(\phi_{j,t},\mathbf{x}_{j,t}\right)$. For example, if one
employs the method proposed by \citet{Blundell2003,Blundell2004},
which accommodates endogenous variables during first-stage estimation,
including raw prices, $p_{j,t}$, in $\mathbf{w}_{j,t}$ can be a
viable choice. However, finding additional exogenous variables that
affect only a consumer's buy-or-not decision is ideal. If one is willing
to assume that $\eta_{j,t}$ is distributed as standard Gaussian,
the classic Heckman correction estimator with instruments can be used,
in which the inverse Mills ratio is added as an additional regressor.
In that case, identification of model parameters is achieved by the
non-linearity of the inverse Mills ratio, and therefore the exclusion
restriction is unnecessary.

\section{\label{sec:Derivation-of-Selection-Correcti}Excursus: Derivation
of the Selection-Correction Estimation Equation for the Logit Demand
Model}

In this section, we provide a two-stage model of consumer choice within
the logit demand frameworks when zero market shares are present. The
logit demand model with two-stage decision process can also lead to
the same estimation equation derived from our proposed CES demand
system, which is presented in Section \ref{sec:Semiparametric-Estimation}.
Although sticking to the logit demand frameworks might be less appealing
because the intensive and extensive margins cannot be distinguished
conceptually, the single-choice assumption can be more adequate in
some contexts. In such contexts, the first-stage decision obtains
an interpretation akin to the consideration set selection in the consideration
set literature.

We show that the estimation equation (\ref{eq:loglineardemand}) can
be derived from the two-stage decision process from the logit demand
model. We consider a representative consumer with a two-stage decision
process. During the first stage, the consumer searches $\mathcal{J}_{t}$,
which includes all possible alternatives. The consumer's consideration
set $\mathcal{J}_{t}^{+}$ is determined from the search. During the
second stage, the consumer encounters the usual discrete choice decision
problem over $\mathcal{J}_{t}^{+}$, that is, purchase the product
that yields the highest utility. Let $\left(\mathbf{w}_{j,t},\eta_{j,t}\right)$
be the variables that affect the first-stage consideration set search,
and $\left(\mathbf{x}_{j,t},\xi_{j,t}\right)$ the variables that
affect the second-stage discrete choice unconstrained utility maximization
problem. Notice that these variables form an analogue of the notations
used in Sections \ref{subsec:The-Exponential-Quality} and \ref{sec:Semiparametric-Estimation}.
The second-stage utility of the consumer is modeled as:
\[
u_{i,j,t}=-\alpha\ln p_{j,t}+\mathbf{x}_{j,t}'\bm{\beta}+\xi_{j,t}+\epsilon_{i,j,t}.
\]
\footnote{By not including the income $y_{i}$, we disregard the ``indirect
utility'' interpretation of the alternative choice utility $u_{i,j,t}$
here. Early literature on the discrete choice consumer demand, which
dates back to \citet{McFadden1974,McFadden1978,McFadden1981,Dubin1984,Anderson1987,Anderson1988,Anderson1992},
stick to the indirect utility interpretation of an alternative utility.
To our understanding, the main intention of interpreting $u_{i,j,t}$
as an indirect utility was to place the discrete choice demand systems
in the context of the Walrasian demand, especially because the price
should not be a direct argument of the Walrasian utility function.
However, discrete choice modeling has gained greater popularity and
has been applied to a much wider context since McFadden's original
works. The ``modern'' approach tend more to interpret $u_{i,j,t}$
as a direct utility of an alternative. Many recent research using
the discrete choice demand estimation framework specify the mean utility
as either linear in $-p_{j,t}$ or in $y_{i}-p_{j,t}$ so the income
$y_{i}$ cancels out over alternatives. See, for example, \citet{Berry1994,Nevo2001}.
\citet{Berry1995} suggest using $-p_{j,t}$ in their homogeneous
coefficients utility specification and $\ln\left(y_{i}-p_{j,t}\right)$
in their random coefficients utility specification. We note that $y_{i}$
does not cancel out over the alternatives in the latter case.}\footnote{In the context of the logit demand models, the utility can be linear,
not log-linear, in prices. If one does not want to interpret the estimated
parameters to be originating from a CES demand system, one can replace
$\ln p_{j,t}$ with $p_{j,t}$.} The representative consumer solves:
\[
\max_{j\in\mathcal{J}_{t}^{+}}\left\{ u_{i,j,t}\right\} .
\]
With the i.i.d. Type-I extreme value assumption on $\epsilon_{i,j,t}$'s,
the individual choice probability becomes: 
\[
\Pr\left(i\rightarrow j|t\right)=\frac{\exp\left(-\alpha\ln p_{j,t}+\mathbf{x}_{j,t}'\bm{\beta}+\xi_{j,t}\right)}{\sum_{k\in\mathcal{J}_{t}^{+}}\exp\left(-\alpha\ln p_{k,t}+\mathbf{x}_{k,t}'\bm{\beta}+\xi_{k,t}\right)}.
\]
$\Pr\left(i\rightarrow j|t\right)$ is the predicted quantity market
share, $\pi_{j,t}$. During estimation, $\pi_{j,t}$ is equated with
the observed market share, $s_{j,t}$. The inversion theorem of \citet{Berry1994,Berry1995}
applies. Again, the only difference is the moment condition; $E\left[\xi_{j,t}|\mathbf{z}_{j,t},d_{j,t}=1\right]$
is not zero and is highly likely to be positive. Thus, a correction
term is needed for the selection of the consideration set, which leads
to the estimation equation: 
\begin{equation}
E\left[\ln\left(\frac{\pi_{j,t}}{\pi_{0,t}}\right)|\mathbf{z}_{j,t},\mathbf{w}_{j,t},d_{j,t}=1\right]=-\alpha\ln p_{j,t}+\mathbf{x}_{j,t}'\bm{\beta}+E\left[\xi_{j,t}|\mathbf{z}_{j,t},\mathbf{w}_{j,t},d_{j,t}=1\right].\label{eq:logitcorrection}
\end{equation}
(\ref{eq:logitcorrection}) coincides with (\ref{eq:loglineardemand}).

\section{\label{sec:Monte-Carlo-Simulations}Monte-Carlo Simulations}

We simulate market data and back out model parameters to examine the
finite-sample performance of the estimator that we proposed in the
previous section. We compare the estimation result using our model
to the estimation result of the logit demand model, which either drops
the sample with zero observed market shares or imputes the zero observed
market shares with a small positive number. The estimator we proposed
in the previous section works well when the model is specified correctly.

We first describe our data-generating process that satisfies the exclusion
restriction. Each market, $t$, has two to five products, with the
exact number of products in each market drawn randomly. The observed
product characteristic vector, $\mathbf{w}_{j,t}$, includes three
continuous components, one discrete component, and three brand dummies.
One of the continuous components is excluded in $\mathbf{x}_{j,t}$.
The first component, $w_{j,t}^{\left(1\right)}$, follows $\mbox{lognormal}\left(0,1\right)$,
the second component, $w_{j,t}^{\left(2\right)}$, follows $\mbox{uniform}\left(1,5\right)$,
the third, $w_{j,t}^{\left(3\right)}$, $\mbox{Poisson}\left(3\right)$,
and the fourth, $w_{j,t}^{\left(4\right)}$, $\mathcal{N}\left(0,1\right)$.
$w_{j,t}^{\left(4\right)}$ is excluded from $\mathbf{x}_{j,t}$.
$\eta_{j,t}|\mathbf{w}_{j,t}$ follows the Type-I extreme value distribution
with mean zero. Two instruments are employed for prices, which are
proxies for cost shocks. Prices, $p_{j,t}$, which is an endogenous
variable, is determined by $p_{j,t}=\psi\left(\mathbf{x}_{j,t},\xi_{j,t}\right)$,
where $\psi$ is some (possibly) nonlinear function that is strictly
monotonic in $\xi_{j,t}$. We specify $\psi$ as: 
\[
\psi\left(\mathbf{z}_{j,t},\xi_{j,t}\right)=2+\frac{1}{50}\left(2z_{j,t}^{\left(1\right)}+4z_{j,t}^{\left(2\right)}+2x_{j,t}^{\left(1\right)}+x_{j,t}^{\left(1\right)}x_{j,t}^{\left(2\right)}-x_{j,t}^{\left(2\right)}x_{j,t}^{\left(3\right)}+5x_{j,t}^{\left(4\right)}+7x_{j,t}^{\left(5\right)}+9x_{j,t}^{\left(6\right)}+8\xi_{j,t}\right).
\]
We intentionally let the influence of the cost proxies, $z_{j,t}^{\left(1\right)}$
and $z_{j,t}^{\left(2\right)}$, to be fairly weak, which reflects
common circumstances in practice. We calibrate the parameters as $\sigma=2$,
$\alpha=1$, $\bm{\beta}=\left(1,-2,1.5,0.3,0.2,0.4\right)'$, $\gamma=\alpha$,
and $\bm{\delta}=\frac{1}{4}\times\left(\bm{\beta},0.1\right)'$,
and market shares are determined by (\ref{eq:individualprob}).

\begin{figure}
\caption{\label{fig:Estimated-Densities-of}Estimated Densities of $\eta_{j,t}|\mathbf{w}_{j,t}$}

\begin{centering}
\includegraphics[width=0.6\paperwidth]{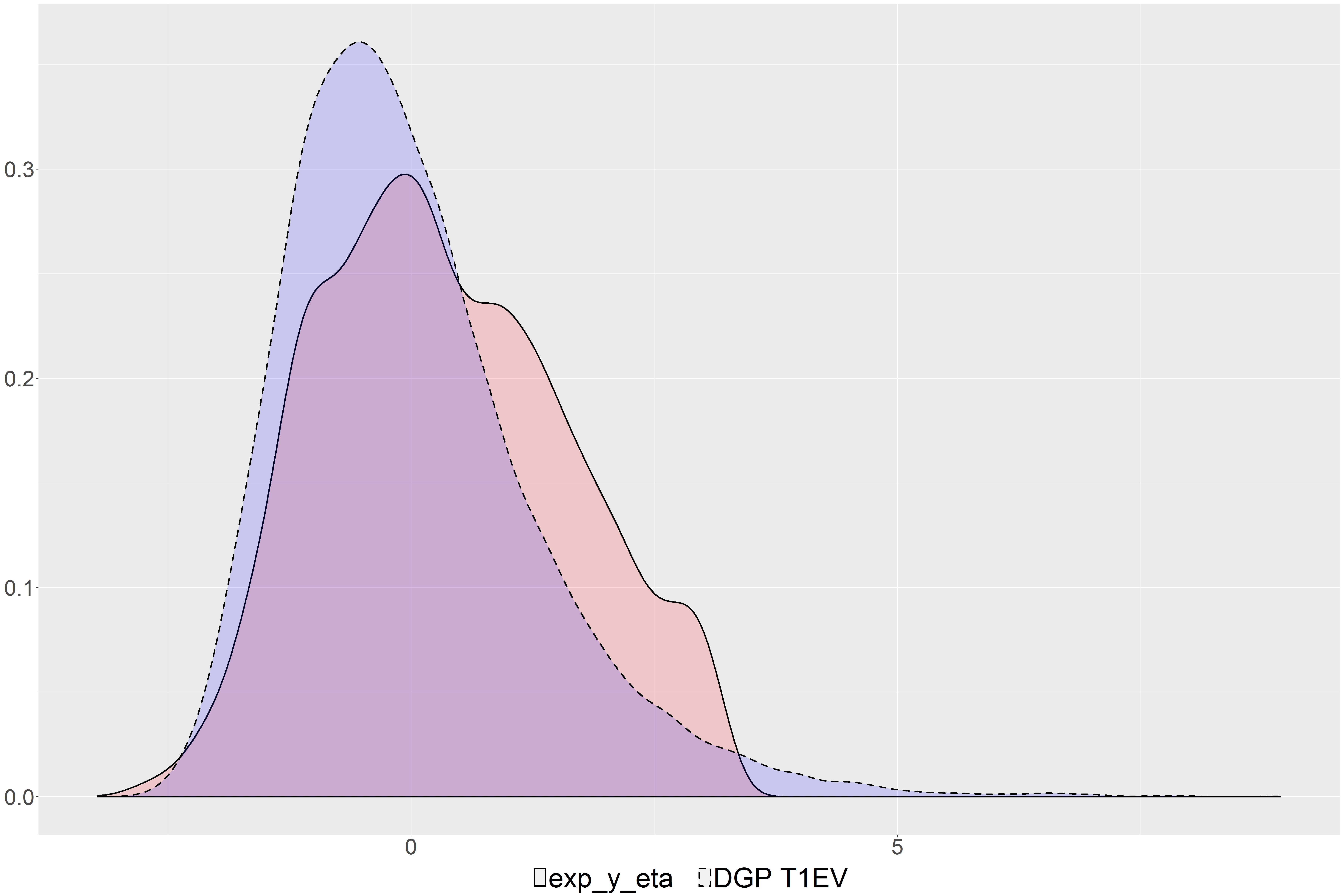}\medskip{}
\par\end{centering}
{\footnotesize{}Note. (i) ``exp\_y\_eta,'' the pink solid density,
is the estimated density of $\eta_{j,t}|\mathbf{w}_{j,t}$ from the
Klein-Spady model. ``DGP T1EV,'' the blue dotted density, is the
Type-I Extreme Value density that is used to generate the data. (ii)
10,000 sample draws are taken and plotted from the estimated density
of the Klein-Spady model and the true Type-I Extreme Value density,
respectively. (iii) The Klein-Spady model identifies the distribution
of unobservables up to location and scale. Thus, we made the location
and scale adjustment. }{\footnotesize\par}
\end{figure}

\begin{table}[!tbph]
\caption{\label{tab:Estimation-Result-of}Estimation Result of the Simulated
Data}

\begin{centering}
\begin{tabular}{cccccc}
\hline 
 &  & (1) & (2) & (3) & (4)\tabularnewline
{\small{}Estimation}  &  & \multicolumn{1}{c}{{\small{}Our Model, K/S}} & \multicolumn{1}{c}{{\small{}Our Model, Heckman}} & {\small{}Logit, Drop 0}  & {\small{}Logit, Impute $10^{-8}$}\tabularnewline
\hline 
\multicolumn{1}{c}{{\small{}Log prices ($-2$)}} &  & {\small{}$\begin{array}{c}
-1.969\\
\left(0.128\right)
\end{array}$}  & {\small{}$\begin{array}{c}
-1.972\\
\left(0.119\right)
\end{array}$}  & {\small{}$\begin{array}{c}
-1.287\\
\left(0.074\right)
\end{array}$}  & {\small{}$\begin{array}{c}
-5.444\\
\left(0.372\right)
\end{array}$}\tabularnewline
\multicolumn{1}{c}{{\small{}$x_{j,t}^{\left(1\right)}$ ($1$)}} &  & {\small{}$\begin{array}{c}
0.807\\
\left(0.014\right)
\end{array}$}  & {\small{}$\begin{array}{c}
1.017\\
\left(0.011\right)
\end{array}$}  & {\small{}$\begin{array}{c}
0.886\\
\left(0.007\right)
\end{array}$}  & {\small{}$\begin{array}{c}
1.536\\
\left(0.041\right)
\end{array}$}\tabularnewline
\multicolumn{1}{c}{{\small{}$x_{j,t}^{\left(2\right)}$ ($-2$)}} &  & {\small{}$\begin{array}{c}
-1.635\\
\left(0.028\right)
\end{array}$}  & {\small{}$\begin{array}{c}
-2.045\\
\left(0.035\right)
\end{array}$}  & {\small{}$\begin{array}{c}
-1.574\\
\left(0.013\right)
\end{array}$}  & {\small{}$\begin{array}{c}
-3.884\\
\left(0.050\right)
\end{array}$}\tabularnewline
\multicolumn{1}{c}{{\small{}$x_{j,t}^{\left(3\right)}$ ($1.5$)}} &  & {\small{}$\begin{array}{c}
1.222\\
\left(0.020\right)
\end{array}$}  & {\small{}$\begin{array}{c}
1.530\\
\left(0.023\right)
\end{array}$}  & {\small{}$\begin{array}{c}
1.339\\
\left(0.008\right)
\end{array}$}  & {\small{}$\begin{array}{c}
1.821\\
\left(0.040\right)
\end{array}$}\tabularnewline
\multicolumn{1}{c}{{\small{}$x_{j,t}^{\left(4\right)}$ ($0.3$)}} &  & {\small{}$\begin{array}{c}
0.251\\
\left(0.051\right)
\end{array}$}  & {\small{}$\begin{array}{c}
0.290\\
\left(0.071\right)
\end{array}$}  & {\small{}$\begin{array}{c}
0.366\\
\left(0.046\right)
\end{array}$}  & {\small{}$\begin{array}{c}
-0.468\\
\left(0.219\right)
\end{array}$}\tabularnewline
\multicolumn{1}{c}{{\small{}$x_{j,t}^{\left(5\right)}$ ($0.2$)}} &  & {\small{}$\begin{array}{c}
0.140\\
\left(0.051\right)
\end{array}$}  & {\small{}$\begin{array}{c}
0.169\\
\left(0.071\right)
\end{array}$}  & {\small{}$\begin{array}{c}
0.266\\
\left(0.046\right)
\end{array}$}  & {\small{}$\begin{array}{c}
-0.673\\
\left(0.220\right)
\end{array}$}\tabularnewline
\multicolumn{1}{c}{{\small{}$x_{j,t}^{\left(6\right)}$ ($0.4$)}} &  & {\small{}$\begin{array}{c}
0.338\\
\left(0.053\right)
\end{array}$}  & {\small{}$\begin{array}{c}
0.435\\
\left(0.073\right)
\end{array}$}  & {\small{}$\begin{array}{c}
0.458\\
\left(0.047\right)
\end{array}$}  & {\small{}$\begin{array}{c}
-0.099\\
\left(0.225\right)
\end{array}$}\tabularnewline
\multicolumn{1}{c}{} &  &  &  &  & \tabularnewline
\multicolumn{1}{c}{{\small{}$D$}} &  & {\small{}$5059$}  & {\small{}$5059$}  & {\small{}$5059$}  & {\small{}$10500$}\tabularnewline
\multicolumn{1}{c}{{\small{}$N$}} &  & {\small{}$10500$}  & {\small{}$10500$}  & {\small{}$5059$}  & {\small{}$10500$}\tabularnewline
\hline 
\end{tabular}\medskip{}
\par\end{centering}
{\footnotesize{}Note. (i) Target values are in parentheses of corresponding
items in the first column. (ii) The Estimation row specifies the method
used during estimation. Column (1) is our proposed estimator, in which
the first-stage propensity score was estimated using the Klein-Spady
estimator. For Column (2), Probit was used for the first-stage propensity
score estimation, and the inverse Mills ratio is added as an additional
regressor during the second stage. Column (3) is the logit estimator
with dropping the samples with zero observed market shares, and Column
(4) is the logit estimator with imputing $10^{-8}$ in place of the
zero observed market shares. (iii) Asymptotic standard error estimates
appear in parentheses. (iv) $D$ is the number of non-censored samples,
and $N$ is the effective sample size.}{\footnotesize\par}
\end{table}

Figure \ref{fig:Estimated-Densities-of} depicts the estimated density
of $\eta_{j,t}|\mathbf{w}_{j,t}$ from the first stage, and compares
it with the distribution used for generating the data. Although the
estimated density does not coincide perfectly with the exact density
of the Type-I extreme value distribution, it preserves the approximate
shape of the distribution. A larger sample is needed for the estimated
densities to fit exactly with the distribution used during data generation.

Table \ref{tab:Estimation-Result-of} shows the estimation results
of the simulated data. The ``Estimation'' row indicates the estimation
method used. Column (1) is the correct quality kernel specification
with our semiparametric estimator, and Column (2) is the correct quality
kernel specification with the classical Heckman correction estimator
assuming Gaussian error term in the first stage. Our estimator is
successful in recovering the true parameters if the model is specified
correctly. The estimator continues to be successful when we estimated
the model using the classical Heckman correction estimator that assumes
the joint normality of the error term distribution. Column (3) is
the logit estimator where we drop the sample with zero observed market
shares, and Column (4) is the logit estimator where we impute small
positive numbers in place of zero observed market shares. Both dropping
zeros and imputing small numbers in place of zeros bias the estimators
substantially. The price coefficient is biased upward when the zero
shares are dropped, whereas it is biased downward when a small number
is imputed in place of zero shares. We also generated and estimated
several other specifications, such as different error term distributions,
functional forms of quality kernels, variables, pricing functions,
etc. For brevity, we do not present all specifications here, but we
note that results and implications presented in this section remain
robust to these alternative specifications. Details on the estimation
procedure appear in Appendix \ref{sec:Implementation-Details}.

\section{\label{sec:Empirical-Example}Empirical Example: Scanner Data with
a Multitude of Zero Shares}

We implement our proposed demand estimation framework using Dominick's
supermarket cola sales scanner data. Data were obtained from the James
M. Kilts Center for Marketing, University of Chicago Booth School
of Business. The data contained weekly pricing and sales information
for the Dominick's chain of stores from 1989 to 1997 for every universal
product code (UPC) level product in 29 product categories. Promotion
statuses and profitability of each unit sold were also recorded in
the data. One shortcoming was that systemic records of product characteristics
were unavailable, which we overcame by choosing cola sales data and
hand-coding the product characteristics.

\subsection{Data}

\begin{figure}
\caption{\label{fig:Histogram-of-Observed}Histogram of the Observed Market
Shares}

\begin{centering}
\includegraphics[width=0.7\paperwidth]{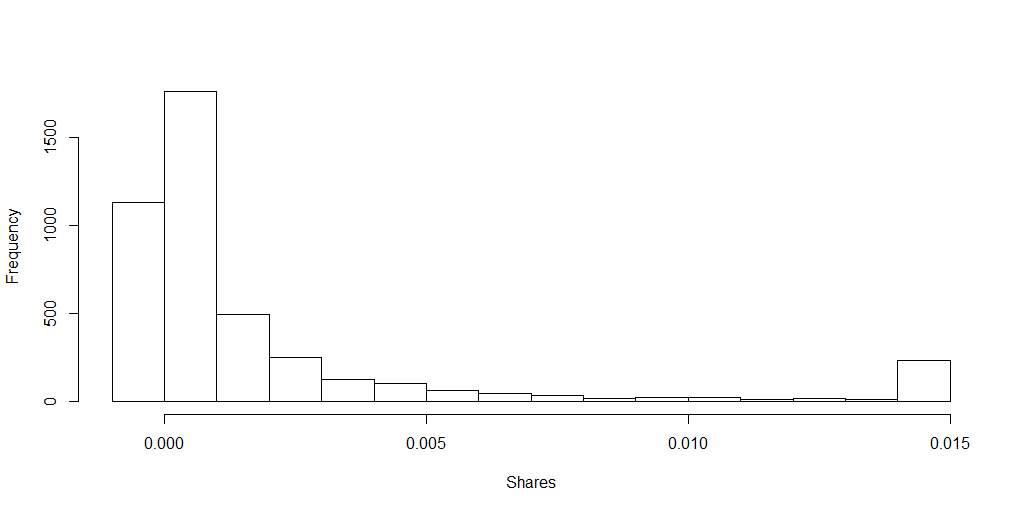}
\par\end{centering}
{\footnotesize{}Note. (i) This figure plots the histogram of the observed
quantity market shares for cola sales of week 391 (03/06/1997 to 03/12/1997)
in Dominick's scanner data. (ii) Sample points larger than 0.015 is
top-coded as 0.015. 216 out of 4356 (4.96\%) sample points are top-coded.
(iii) 1130 out of 4356 samples (25.94\%) have zero market shares.}{\footnotesize\par}
\end{figure}

We chose Dominick's data because they were ideal for illustrating
the application of our framework for two reasons. First, Dominick's
data contained information on which products were displayed on the
shelves, even if a product did not sell in the corresponding week
and store. This feature was necessary because we wanted the exact
information on products that were in a consumer's consideration set
but were not chosen. Presented in Figure \ref{fig:Histogram-of-Observed},
approximately one-fourth of observations exhibited zero observed market
shares. Second, Dominick's data contained information on average profit
per unit sold. Combined with price data, we could back out the average
cost per unit. Cost information is useful because an ideal instrument
for prices when estimating consumer demand should proxy cost shocks.
We avoided constructing instruments using indirect proxies for cost,
which has been a major difficulty in demand estimation literature.

We focus on cola sales for several reasons. First, the cola market
is a typical market of product differentiation, in which many brands
with disparate tastes and packages competes. Among them, Coke and
Pepsi, the two prominent brands, take the majority of market shares.
Second, product characteristics were not coded separately in Dominick's
data, but only category information such as ``soft drinks'' or ``bottled
juices.'' We had to extract the information from product descriptions
truncated at 30 characters, for which cola was ideal because it had
clearly labeled product characteristics. Finally, companies producing
cola, and product characteristics of cola, have not changed much during
the past few decades. Coke and Pepsi have been two leaders in the
market. Diet, cherry-flavored, and caffeine-free colas are still sold
in the market with considerable market shares in 2016, and in 1996.
This feature made our analysis convenient, and the implications of
analysis more realistic. In Appendices \ref{subsec:Laundry-Detergent-Data}
and \ref{subsec:Laundry-Detergent-Demand}, we present estimation
results for laundry detergent demand as a robustness check.

Dominick's data covered 100 chain stores in the Chicago area for 400
weeks, from September 1989 to May 1997. We chose the cross-section
of week 391, which is the second week of March 1997. We used the cross-section
data of a week because demand for soft drinks fluctuates in weeks
with holidays or events such as the Super Bowl, and varies considerably
by season. Therefore, we chose a week in March without any close holidays.
As Dominick's experimented with prices across chain stores for the
same product during the same week, we still have sufficient price
variations after choosing a cross-section of data. Even after restricting
the sample to a cross-section of one week, the sample size was as
large as 4,300. We present summary statistics in Table \ref{tab:Summary-Statistics}.

We define individual products and markets naturally. An individual
product was defined by its UPC, and a market by a store-week pair.
This was the finest manner of defining a product and market that the
data allowed, which resulted in a multitude of zero observed market
shares. Illustrated in Figure \ref{fig:Histogram-of-Observed}, approximately
one-fourth of products that were displayed on shelves did not sell.

\begin{table}[!tbph]
\caption{\label{tab:Summary-Statistics}Descriptive Statistics}

\begin{centering}
{\small{}}\subfloat[Summary of the Product Characteristics of the Sample]{{\small{}}{\small\par}
\centering{}{\small{}}%
\begin{tabular}{lccc}
\hline 
 & {\small{}Frequency} & {\small{}Mean} & {\small{}Std}\tabularnewline
\hline 
\hline 
{\small{}Diet} & {\small{}2163} & {\small{}0.497} & {\small{}0.500}\tabularnewline
{\small{}Caffeine Free} & {\small{}1085} & {\small{}0.249} & {\small{}0.433}\tabularnewline
{\small{}Cherry} & {\small{}151} & {\small{}0.035} & {\small{}0.183}\tabularnewline
{\small{}Coke} & {\small{}365} & {\small{}0.084} & {\small{}0.277}\tabularnewline
{\small{}Pepsi} & {\small{}2644} & {\small{}0.607} & {\small{}0.488}\tabularnewline
{\small{}Promo} & {\small{}1751} & {\small{}0.402} & {\small{}0.490}\tabularnewline
{\small{}Bottle Size} & {\small{}-} & {\small{}26.592} & {\small{}29.696}\tabularnewline
{\small{}\# Bottles per Bundle} & {\small{}-} & {\small{}12.436} & {\small{}9.667}\tabularnewline
{\small{}\# Stores} & {\small{}73} & {\small{}-} & {\small{}-}\tabularnewline
{\small{}Uncensored Obs ($D$)} & {\small{}3226} & {\small{}-} & {\small{}-}\tabularnewline
{\small{}Sample Size ($N$)} & {\small{}4356} & {\small{}-} & {\small{}-}\tabularnewline
\hline 
\end{tabular}{\small\par}}{\small\par}
\par\end{centering}

\begin{centering}
{\small{}}\subfloat[Per-ounce Price, Cost, Profitability, and Market Shares of Products
in the Full Sample]{{\small{}}{\small\par}
\centering{}{\small{}}%
\begin{tabular}{lccccc}
\hline 
 & {\small{}Mean} & {\small{}Median} & {\small{}Std} & {\small{}Min} & {\small{}Max}\tabularnewline
\hline 
\hline 
{\small{}Per-ounce Prices (\$)} & {\small{}0.020} & {\small{}0.024} & {\small{}0.014} & {\small{}0} & {\small{}0.042}\tabularnewline
{\small{}Per-ounce Cost (\$)} & {\small{}0.014} & {\small{}0.017} & {\small{}0.010} & {\small{}0} & {\small{}0.028}\tabularnewline
{\small{}Profitability (\%)} & {\small{}20.470} & {\small{}28.380} & {\small{}16.020} & {\small{}$-$98.550} & {\small{}58.620}\tabularnewline
{\small{}Shares (\%)} & {\small{}0.586} & {\small{}0.038} & {\small{}2.879} & {\small{}0} & {\small{}42.967}\tabularnewline
\hline 
\end{tabular}{\small\par}}{\small\par}
\par\end{centering}

\begin{centering}
{\small{}}\subfloat[Per-ounce Price, Cost, Profitability, and Market Shares of Products
in the Noncensored Sample]{{\small{}}{\small\par}
\centering{}{\small{}}%
\begin{tabular}{lccccc}
\hline 
 & {\small{}Mean} & {\small{}Median} & {\small{}Std} & {\small{}Min} & {\small{}Max}\tabularnewline
\hline 
\hline 
{\small{}Per-ounce Prices (\$)} & {\small{}0.027} & {\small{}0.026} & {\small{}0.008} & {\small{}0.005} & {\small{}0.042}\tabularnewline
{\small{}Per-ounce Cost (\$)} & {\small{}0.019} & {\small{}0.019} & {\small{}0.006} & {\small{}0.003} & {\small{}0.028}\tabularnewline
{\small{}Profitability (\%)} & {\small{}27.640} & {\small{}29.180} & {\small{}12.499} & {\small{}$-$98.550} & {\small{}58.620}\tabularnewline
{\small{}Shares (\%)} & {\small{}0.791} & {\small{}0.084} & {\small{}3.321} & {\small{}0.001} & {\small{}42.967}\tabularnewline
\hline 
\end{tabular}{\small\par}}{\small\par}
\par\end{centering}
{\footnotesize{}Note. (i) Data are the cross-section of week 391 (03/06/1997
to 03/12/1997) in Dominick's scanner data. (ii) Dominick's recorded
the price and cost as zero if sales of a product were zero in a corresponding
week. The mean and median of price and cost in Table \ref{tab:Summary-Statistics}-(b)
were calculated including those zeros. }{\footnotesize\par}
\end{table}

We converted package prices and costs to per-ounce prices and costs.
Dominick's did not record the price and cost of the week if sales
of a product were zero in a corresponding week. Therefore, we could
not include prices in $\mathbf{w}_{j,t}$, and proceeded only with
other exogenous variables during first-stage estimation. When estimating
the logit model while substituting the zero observed market shares
with small numbers, we imputed missing prices and costs using other
chain stores' prices and profits with the same product and promotion
status. We had to compute market shares of outside options for both
our model and the logit demand model.\footnote{Although including the numeraire in a consumer's consideration set
was unnecessary in our model, we included it because we wanted to
compare estimation results of our model with those of the logit model
using the same setup.} When estimating market size, we assumed that an average person consumed
100 ounces of soft drinks a week,\footnote{On average, Americans consume about 45 gallons of soft drinks a year.
Source: \href{http://adage.com/article/news/consumers-drink-soft-drinks-water-beer/228422/}{http://adage.com/article/news/consumers-drink-soft-drinks-water-beer/228422/}.} and computed the size of the market using daily customer count data
for each store in the chain.

\subsection{Estimation, Result, and Discussion}

We estimate our model using the method proposed in Section \ref{sec:Semiparametric-Estimation}.
We also estimate the model correcting for a consumer's consideration
set selection using the Probit as a first-stage estimator, with the
\citet{Powell2001} estimator and the simple Heckman selection correction
estimator during the second stage. The simple Heckman estimator was
implemented using the inverse Mills ratio as an additional regressor
as usual. As a benchmark, we estimated the homogeneous logit model
of demand, with different ways of handling the zero observed market
shares: (i) dropping samples with zero observed market shares, and
(ii) substituting zero observed market shares with small numbers.
We also used the log of prices in the logit model to compare the magnitudes
of coefficients. Mentioned previously, using the log of prices instead
of raw prices represents a scale adjustment in the utility specification
of the logit demand model.

We estimated two models with different specifications. In the baseline
model (Model 1), $\mathbf{x}_{j,t}$ includes several product characteristics:
bottle size, number of bottles per bundle, diet, caffeine-free, cherry
flavor, Coke/Pepsi brand dummies. As an instrument of the per-ounce
price, we used the per-ounce cost calculated from the profitability
variable. For Model 1, we excluded promotion status from $\mathbf{x}_{j,t}$,
and use it as a variable that satisfies the exclusion restriction.
The exclusion assumption in this case reflects the informational hypothesis:
promotions affect only consumers' information about a consideration
set, not the level of utility associated with consuming a certain
product. For Model 2, we included the promotion statuses in $\mathbf{x}_{j,t}$,
and used store-level demographics for variables included in $\mathbf{w}_{j,t}$
that were not included in $\mathbf{x}_{j,t}$: \% Blacks and Hispanics,
\% college graduates, and log of the median income. The exclusion
assumption of these variables reflects the preferential hypothesis
of the extensive margin: a consumer who never buys a certain product
will not become an inframarginal consumer regardless of other product
characteristics. 

\begin{figure}[!tbph]
\caption{\label{fig:Estimated-Distribution-of}Estimated Densities of $\eta_{j,t}|\mathbf{w}_{j,t}$}

\begin{centering}
\includegraphics[width=0.6\paperwidth]{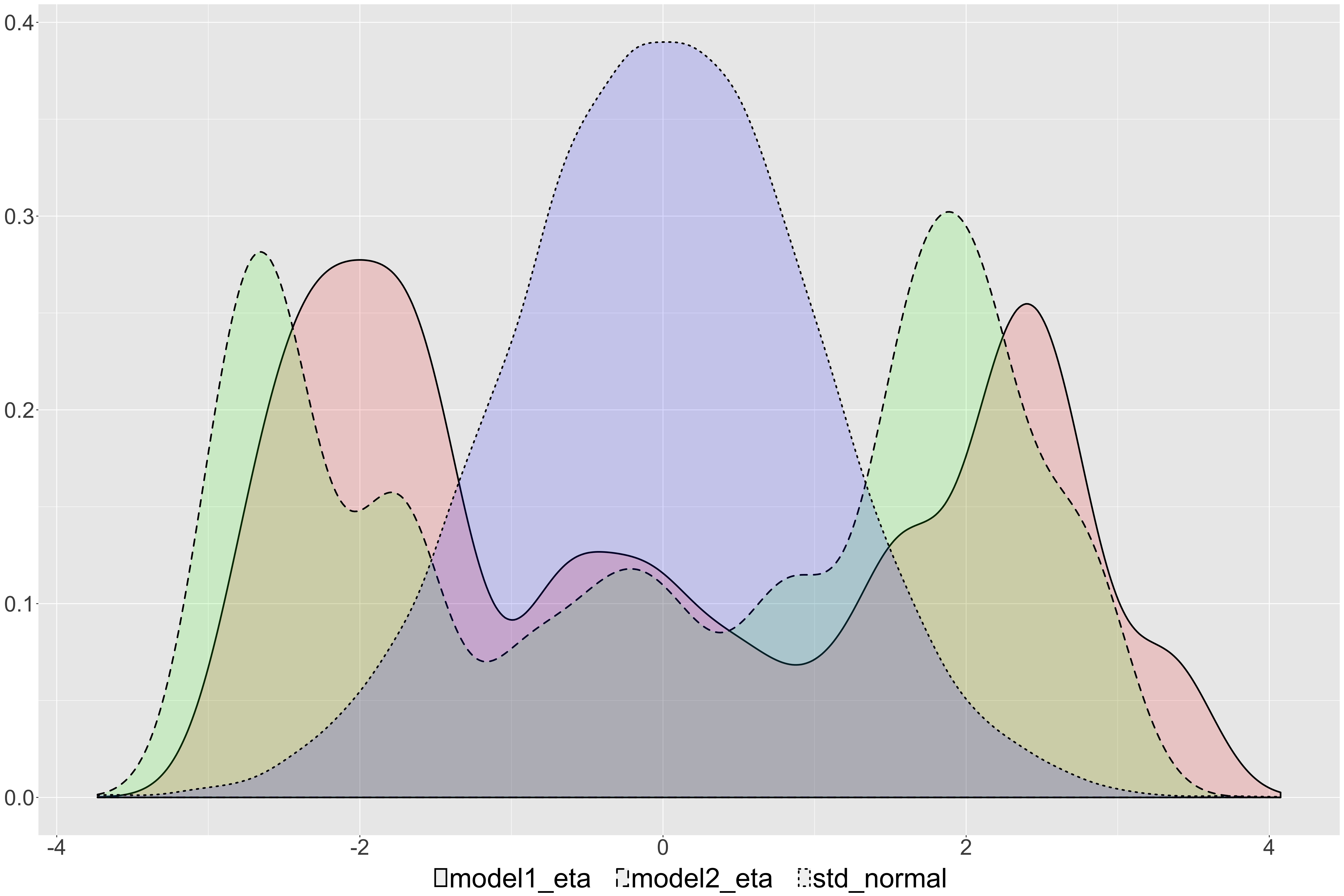}\medskip{}
\par\end{centering}
{\footnotesize{}Note. (i) ``model1\_eta,'' the pink solid density,
is the estimated density of $\eta_{j,t}|\mathbf{w}_{j,t}$ from Model
1. ``model2\_eta,'' the green dotted density, is the estimated density
of $\eta_{j,t}|\mathbf{w}_{j,t}$ from Model 2. ``std\_normal,''
the blue dotted density, is the standard normal density plotted for
benchmark. (ii) The Klein-Spady model identifies the distribution
of unobservables up to location and scale, and thus we made a location
and scale adjustment of $E\left[\eta_{j,t}|\mathbf{w}_{j,t}\right]=0$
and $Var\left(\eta_{j,t}|\mathbf{w}_{j,t}\right)=4$. (iii) For Model
1 and 2, 10,000 sample draws were taken from the density estimates
of the Klein-Spady model, and the density of the drawn sample was
then plotted. (iv) The density of 10,000 sample draws from the standard
Gaussian distribution is plotted for comparison.}{\footnotesize\par}
\end{figure}

\begin{table}[!tbph]
\caption{\label{tab:First-stage-Parameter-Estimates}First-stage Parameter
Estimates $\hat{\bm{\delta}}$ }

\begin{centering}
\begin{tabular}{cccccc}
\hline 
 & \multicolumn{2}{c}{{\footnotesize{}Model 1}} &  & \multicolumn{2}{c}{{\footnotesize{}Model 2}}\tabularnewline
\cline{2-3} \cline{5-6} 
 & {\footnotesize{}Probit}  & {\footnotesize{}Klein-Spady} &  & {\footnotesize{}Probit}  & {\footnotesize{}Klein-Spady} \tabularnewline
{\footnotesize{}$\mathbf{w}_{j,t}$}  & {\footnotesize{}(1)} & {\footnotesize{}(2)} &  & {\footnotesize{}(3)} & {\footnotesize{}(4)}\tabularnewline
\hline 
{\footnotesize{}Bottle Size}  & {\footnotesize{}$\begin{array}{c}
1\\
\left(0.195\right)
\end{array}$}  & {\footnotesize{}$\begin{array}{c}
1\\
\left(-\right)
\end{array}$} &  & {\footnotesize{}$\begin{array}{c}
1\\
\left(0.196\right)
\end{array}$}  & {\footnotesize{}$\begin{array}{c}
1\\
\left(-\right)
\end{array}$} \tabularnewline
{\footnotesize{}\# Bottles per Bundle}  & {\footnotesize{}$\begin{array}{c}
-1.431\\
\left(0.329\right)
\end{array}$}  & {\footnotesize{}$\begin{array}{c}
-91.991\\
\left(3.780\right)
\end{array}$} &  & {\footnotesize{}$\begin{array}{c}
-1.453\\
\left(0.330\right)
\end{array}$}  & {\footnotesize{}$\begin{array}{c}
-133.500\\
\left(12.535\right)
\end{array}$} \tabularnewline
{\footnotesize{}Diet}  & {\footnotesize{}$\begin{array}{c}
18.342\\
\left(4.818\right)
\end{array}$}  & {\footnotesize{}$\begin{array}{c}
-277.300\\
\left(12.238\right)
\end{array}$} &  & {\footnotesize{}$\begin{array}{c}
18.222\\
\left(4.822\right)
\end{array}$}  & {\footnotesize{}$\begin{array}{c}
-515.187\\
\left(50.981\right)
\end{array}$} \tabularnewline
{\footnotesize{}Caffeine Free}  & {\footnotesize{}$\begin{array}{c}
-13.619\\
\left(6.158\right)
\end{array}$}  & {\footnotesize{}$\begin{array}{c}
149.733\\
\left(10.108\right)
\end{array}$} &  & {\footnotesize{}$\begin{array}{c}
-13.663\\
\left(6.171\right)
\end{array}$}  & {\footnotesize{}$\begin{array}{c}
118.946\\
\left(4.446\right)
\end{array}$} \tabularnewline
{\footnotesize{}Cherry}  & {\footnotesize{}$\begin{array}{c}
-53.619\\
\left(13.821\right)
\end{array}$}  & {\footnotesize{}$\begin{array}{c}
19.536\\
\left(4.700\right)
\end{array}$} &  & {\footnotesize{}$\begin{array}{c}
-53.582\\
\left(13.805\right)
\end{array}$}  & {\footnotesize{}$\begin{array}{c}
-22.960\\
\left(4.620\right)
\end{array}$} \tabularnewline
{\footnotesize{}Coke}  & {\footnotesize{}$\begin{array}{c}
-41.949\\
\left(10.192\right)
\end{array}$}  & {\footnotesize{}$\begin{array}{c}
-20.901\\
\left(3.607\right)
\end{array}$} &  & {\footnotesize{}$\begin{array}{c}
-41.915\\
\left(10.209\right)
\end{array}$}  & {\footnotesize{}$\begin{array}{c}
91.351\\
\left(4.797\right)
\end{array}$} \tabularnewline
{\footnotesize{}Pepsi}  & {\footnotesize{}$\begin{array}{c}
79.989\\
\left(5.917\right)
\end{array}$}  & {\footnotesize{}$\begin{array}{c}
-70.255\\
\left(2.492\right)
\end{array}$} &  & {\footnotesize{}$\begin{array}{c}
79.904\\
\left(5.910\right)
\end{array}$}  & {\footnotesize{}$\begin{array}{c}
-170.295\\
\left(8.323\right)
\end{array}$} \tabularnewline
{\footnotesize{}Promo}  & {\footnotesize{}$\begin{array}{c}
135.449\\
\left(8.679\right)
\end{array}$}  & {\footnotesize{}$\begin{array}{c}
396.440\\
\left(18.234\right)
\end{array}$} &  & {\footnotesize{}$\begin{array}{c}
135.534\\
\left(8.686\right)
\end{array}$}  & {\footnotesize{}$\begin{array}{c}
601.361\\
\left(53.123\right)
\end{array}$} \tabularnewline
{\footnotesize{}\% Blacks and Hispanics}  & {\footnotesize{}-}  & {\footnotesize{}-} &  & {\footnotesize{}$\begin{array}{c}
-29.891\\
\left(21.430\right)
\end{array}$}  & {\footnotesize{}$\begin{array}{c}
-6.454\\
\left(13.069\right)
\end{array}$} \tabularnewline
{\footnotesize{}\% College Graduates}  & {\footnotesize{}-}  & {\footnotesize{}-} &  & {\footnotesize{}$\begin{array}{c}
3.850\\
\left(20.978\right)
\end{array}$}  & {\footnotesize{}$\begin{array}{c}
4.550\\
\left(12.624\right)
\end{array}$} \tabularnewline
{\footnotesize{}Log Median Income}  & {\footnotesize{}-}  & {\footnotesize{}-} &  & {\footnotesize{}$\begin{array}{c}
-16.888\\
\left(93.441\right)
\end{array}$}  & {\footnotesize{}$\begin{array}{c}
26.399\\
\left(43.820\right)
\end{array}$} \tabularnewline
 &  &  &  &  & \tabularnewline
{\footnotesize{}$D$}  & {\footnotesize{}$3226$}  & {\footnotesize{}$3226$} &  & {\footnotesize{}$3226$}  & {\footnotesize{}$3226$} \tabularnewline
{\footnotesize{}$N$}  & {\footnotesize{}$4356$}  & {\footnotesize{}$4356$} &  & {\footnotesize{}$4356$}  & {\footnotesize{}$4356$} \tabularnewline
\hline 
\end{tabular}\medskip{}
\par\end{centering}
{\footnotesize{}Note. (i) $D$ is the number of non-zero market share
observations, and $N$ is the sample size. (ii) Asymptotic standard
error estimates appear in parentheses. (iii) The unit of bottle size
is liquid ounces. (iv) We normalized the coefficients of the Bottle
Size variable to one. }{\footnotesize\par}
\end{table}

\begin{sidewaystable}[H]
\caption{\label{tab:Parameter-Estimates}Implied Own Price Elasticities and
Second-stage Parameter Estimates $\left(\hat{\sigma},\hat{\bm{\beta}}\right)$}

\begin{centering}
{\footnotesize{}}%
\begin{tabular}{ccccccccccccc}
\hline 
 & \multicolumn{2}{c}{{\footnotesize{}Our Model, K/S }} &  & \multicolumn{2}{c}{{\footnotesize{}Our Model, Probit}} &  & \multicolumn{2}{c}{{\footnotesize{}Heckman Correction}} &  & \multicolumn{3}{c}{{\footnotesize{}Logit Model}}\tabularnewline
\cline{2-3} \cline{5-6} \cline{8-9} \cline{11-13} 
 & {\footnotesize{}Model 1 } & {\footnotesize{}Model 2 } &  & {\footnotesize{}Model 1 } & {\footnotesize{}Model 2 } &  & {\footnotesize{}Model 1 } & {\footnotesize{}Model 2 } &  & {\footnotesize{}Drop 0 } & {\footnotesize{}$10^{-8}$ } & {\footnotesize{}$10^{-4}$}\tabularnewline
 & {\footnotesize{}(1)} & {\footnotesize{}(2)} &  & {\footnotesize{}(3)} & {\footnotesize{}(4)} &  & {\footnotesize{}(5)} & {\footnotesize{}(6)} &  & {\footnotesize{}(7)} & {\footnotesize{}(8)} & {\footnotesize{}(9)}\tabularnewline
\hline 
\hline 
{\footnotesize{}Mean Price Elasticity (CES)} & {\footnotesize{}$-1.326$} & {\footnotesize{}$-1.369$} &  & {\footnotesize{}$-1.138$} & {\footnotesize{}$-1.139$} &  & {\footnotesize{}$-1.299$} & {\footnotesize{}$-1.290$} &  & {\footnotesize{}-} & {\footnotesize{}-} & {\footnotesize{}-}\tabularnewline
{\footnotesize{}Mean Price Elasticity (Logit)} & {\footnotesize{}$-1.324$} & {\footnotesize{}$-1.367$} &  & {\footnotesize{}$-1.133$} & {\footnotesize{}$-1.135$} &  & {\footnotesize{}$-1.297$} & {\footnotesize{}$-1.287$} &  & {\footnotesize{}$4.777$} & {\footnotesize{}$7.703$} & {\footnotesize{}$4.726$}\tabularnewline
\hline 
{\footnotesize{}Log Price ($-\sigma$ / $-\alpha$) } & {\footnotesize{}$\begin{array}{c}
-1.331\\
\left(0.041\right)
\end{array}$ } & {\footnotesize{}$\begin{array}{c}
-1.375\\
\left(0.040\right)
\end{array}$ } &  & {\footnotesize{}$\begin{array}{c}
-1.140\\
\left(0.058\right)
\end{array}$ } & {\footnotesize{}$\begin{array}{c}
-1.142\\
\left(0.061\right)
\end{array}$ } &  & {\footnotesize{}$\begin{array}{c}
-1.304\\
\left(0.040\right)
\end{array}$ } & {\footnotesize{}$\begin{array}{c}
-1.295\\
\left(0.041\right)
\end{array}$ } &  & {\footnotesize{}$\begin{array}{c}
4.805\\
\left(0.099\right)
\end{array}$} & {\footnotesize{}$\begin{array}{c}
7.748\\
\left(0.158\right)
\end{array}$} & {\footnotesize{}$\begin{array}{c}
4.754\\
\left(0.074\right)
\end{array}$}\tabularnewline
{\footnotesize{}Bottle Size } & {\footnotesize{}$\begin{array}{c}
0.011\\
\left(0.002\right)
\end{array}$ } & {\footnotesize{}$\begin{array}{c}
0.009\\
\left(0.001\right)
\end{array}$ } &  & {\footnotesize{}$\begin{array}{c}
0.020\\
\left(0.010\right)
\end{array}$ } & {\footnotesize{}$\begin{array}{c}
0.021\\
\left(0.019\right)
\end{array}$ } &  & {\footnotesize{}$\begin{array}{c}
0.013\\
\left(0.001\right)
\end{array}$ } & {\footnotesize{}$\begin{array}{c}
0.011\\
\left(0.002\right)
\end{array}$ } &  & {\footnotesize{}$\begin{array}{c}
0.121\\
\left(0.004\right)
\end{array}$} & {\footnotesize{}$\begin{array}{c}
0.219\\
\left(0.007\right)
\end{array}$} & {\footnotesize{}$\begin{array}{c}
0.118\\
\left(0.003\right)
\end{array}$}\tabularnewline
{\footnotesize{}\# Bottles per Bundle } & {\footnotesize{}$\begin{array}{c}
0.167\\
\left(0.010\right)
\end{array}$ } & {\footnotesize{}$\begin{array}{c}
0.060\\
\left(0.025\right)
\end{array}$ } &  & {\footnotesize{}$\begin{array}{c}
0.176\\
\left(0.019\right)
\end{array}$ } & {\footnotesize{}$\begin{array}{c}
0.171\\
\left(0.010\right)
\end{array}$ } &  & {\footnotesize{}$\begin{array}{c}
0.130\\
\left(0.003\right)
\end{array}$ } & {\footnotesize{}$\begin{array}{c}
0.136\\
\left(0.006\right)
\end{array}$ } &  & {\footnotesize{}$\begin{array}{c}
0.385\\
\left(0.012\right)
\end{array}$} & {\footnotesize{}$\begin{array}{c}
0.589\\
\left(0.020\right)
\end{array}$} & {\footnotesize{}$\begin{array}{c}
0.357\\
\left(0.009\right)
\end{array}$}\tabularnewline
{\footnotesize{}Diet } & {\footnotesize{}$\begin{array}{c}
0.361\\
\left(0.072\right)
\end{array}$ } & {\footnotesize{}$\begin{array}{c}
0.207\\
\left(0.124\right)
\end{array}$ } &  & {\footnotesize{}$\begin{array}{c}
-0.350\\
\left(0.160\right)
\end{array}$ } & {\footnotesize{}$\begin{array}{c}
-0.316\\
\left(0.112\right)
\end{array}$ } &  & {\footnotesize{}$\begin{array}{c}
-0.186\\
\left(0.045\right)
\end{array}$ } & {\footnotesize{}$\begin{array}{c}
-0.219\\
\left(0.056\right)
\end{array}$ } &  & {\footnotesize{}$\begin{array}{c}
0.943\\
\left(0.118\right)
\end{array}$} & {\footnotesize{}$\begin{array}{c}
2.241\\
\left(0.206\right)
\end{array}$} & {\footnotesize{}$\begin{array}{c}
0.940\\
\left(0.097\right)
\end{array}$}\tabularnewline
{\footnotesize{}Caffeine Free } & {\footnotesize{}$\begin{array}{c}
-1.369\\
\left(0.058\right)
\end{array}$ } & {\footnotesize{}$\begin{array}{c}
-1.198\\
\left(0.054\right)
\end{array}$ } &  & {\footnotesize{}$\begin{array}{c}
-1.104\\
\left(0.178\right)
\end{array}$ } & {\footnotesize{}$\begin{array}{c}
-1.104\\
\left(0.116\right)
\end{array}$ } &  & {\footnotesize{}$\begin{array}{c}
-1.072\\
\left(0.056\right)
\end{array}$ } & {\footnotesize{}$\begin{array}{c}
-1.052\\
\left(0.063\right)
\end{array}$ } &  & {\footnotesize{}$\begin{array}{c}
-2.157\\
\left(0.131\right)
\end{array}$} & {\footnotesize{}$\begin{array}{c}
-2.924\\
\left(0.229\right)
\end{array}$} & {\footnotesize{}$\begin{array}{c}
-1.859\\
\left(0.108\right)
\end{array}$}\tabularnewline
{\footnotesize{}Cherry } & {\footnotesize{}$\begin{array}{c}
-2.990\\
\left(0.123\right)
\end{array}$ } & {\footnotesize{}$\begin{array}{c}
-2.927\\
\left(0.132\right)
\end{array}$ } &  & {\footnotesize{}$\begin{array}{c}
-3.086\\
\left(0.835\right)
\end{array}$ } & {\footnotesize{}$\begin{array}{c}
-3.085\\
\left(1.071\right)
\end{array}$ } &  & {\footnotesize{}$\begin{array}{c}
-2.232\\
\left(0.176\right)
\end{array}$ } & {\footnotesize{}$\begin{array}{c}
-2.139\\
\left(0.185\right)
\end{array}$ } &  & {\footnotesize{}$\begin{array}{c}
-3.452\\
\left(0.343\right)
\end{array}$} & {\footnotesize{}$\begin{array}{c}
-5.367\\
\left(0.535\right)
\end{array}$} & {\footnotesize{}$\begin{array}{c}
-2.915\\
\left(0.252\right)
\end{array}$}\tabularnewline
{\footnotesize{}Coke } & {\footnotesize{}$\begin{array}{c}
-0.375\\
\left(0.103\right)
\end{array}$ } & {\footnotesize{}$\begin{array}{c}
-0.181\\
\left(0.115\right)
\end{array}$ } &  & {\footnotesize{}$\begin{array}{c}
-0.828\\
\left(0.255\right)
\end{array}$ } & {\footnotesize{}$\begin{array}{c}
-0.729\\
\left(0.549\right)
\end{array}$ } &  & {\footnotesize{}$\begin{array}{c}
0.063\\
\left(0.102\right)
\end{array}$ } & {\footnotesize{}$\begin{array}{c}
0.071\\
\left(0.113\right)
\end{array}$ } &  & {\footnotesize{}$\begin{array}{c}
0.593\\
\left(0.252\right)
\end{array}$} & {\footnotesize{}$\begin{array}{c}
1.151\\
\left(0.393\right)
\end{array}$} & {\footnotesize{}$\begin{array}{c}
0.472\\
\left(0.186\right)
\end{array}$}\tabularnewline
{\footnotesize{}Pepsi } & {\footnotesize{}$\begin{array}{c}
1.644\\
\left(0.069\right)
\end{array}$ } & {\footnotesize{}$\begin{array}{c}
1.699\\
\left(0.066\right)
\end{array}$ } &  & {\footnotesize{}$\begin{array}{c}
0.977\\
\left(0.191\right)
\end{array}$ } & {\footnotesize{}$\begin{array}{c}
1.120\\
\left(0.962\right)
\end{array}$ } &  & {\footnotesize{}$\begin{array}{c}
1.068\\
\left(0.062\right)
\end{array}$ } & {\footnotesize{}$\begin{array}{c}
0.868\\
\left(0.191\right)
\end{array}$ } &  & {\footnotesize{}$\begin{array}{c}
4.398\\
\left(0.172\right)
\end{array}$} & {\footnotesize{}$\begin{array}{c}
9.090\\
\left(0.269\right)
\end{array}$} & {\footnotesize{}$\begin{array}{c}
4.374\\
\left(0.127\right)
\end{array}$}\tabularnewline
{\footnotesize{}Promo } & {\footnotesize{}- } & {\footnotesize{}$\begin{array}{c}
0.671\\
\left(0.130\right)
\end{array}$ } &  & {\footnotesize{}- } & {\footnotesize{}$\begin{array}{c}
0.168\\
\left(1.446\right)
\end{array}$ } &  & {\footnotesize{}- } & {\footnotesize{}$\begin{array}{c}
-0.260\\
\left(0.233\right)
\end{array}$ } &  & {\footnotesize{}-} & {\footnotesize{}-} & {\footnotesize{}-}\tabularnewline
 &  &  &  &  &  &  &  &  &  &  &  & \tabularnewline
{\footnotesize{}$D$ } & {\footnotesize{}$3226$ } & {\footnotesize{}$3226$ } &  & {\footnotesize{}$3226$ } & {\footnotesize{}$3226$ } &  & {\footnotesize{}$3226$ } & {\footnotesize{}$3226$ } &  & {\footnotesize{}$3226$ } & {\footnotesize{}$4090$ } & {\footnotesize{}$4090$}\tabularnewline
{\footnotesize{}$N$ } & {\footnotesize{}$4356$ } & {\footnotesize{}$4356$ } &  & {\footnotesize{}$4356$ } & {\footnotesize{}$4356$ } &  & {\footnotesize{}$4356$ } & {\footnotesize{}$4356$ } &  & {\footnotesize{}$3226$ } & {\footnotesize{}$4090$ } & {\footnotesize{}$4090$}\tabularnewline
\hline 
\end{tabular}\medskip{}
\par\end{centering}
{\footnotesize{}Note. (i) For columns ``Our Model, K/S'' and ``Our
Model, Probit,'' results from the Klein-Spady and Probit estimators
in Table \ref{tab:First-stage-Parameter-Estimates} were used for
the first-stage estimator, respectively. Then, the pairwise differenced
weighted instrumental variable estimator was used during the second
stage. For the ``Heckman Correction'' columns, Probit was used during
the first stage, and Heckman's selection correction estimator with
the inverse Mills ratio as an additional regressor was used during
the second stage. (ii) Row ``Mean Price Elasticity (CES)'' is the
mean of the implied Marshallian own price elasticity over the sample
for the CES demand system. Row ``Mean Price Elasticity (Logit)''
is the mean of the implied own price elasticity over the sample for
the Logit demand system when the alternative utility is log-linear
in prices. (iii) $D$ is the number of non-zero market share observations,
and $N$ is the effective sample size. (iv) Asymptotic standard error
estimates are in parentheses. (v) The unit of bottle size is liquid
ounces. (vi) Because Dominick's did not record the price and cost
when sales were zero, when estimating the $10^{-8}$ and $10^{-4}$
columns, we used average prices and costs of the same product with
the same promotion statuses from other stores.}{\footnotesize\par}
\end{sidewaystable}

The first-stage parameter estimation result for $\hat{\bm{\delta}}$
is shown in Table \ref{tab:First-stage-Parameter-Estimates}. Model
1 is the baseline model, with promotion statuses as excluded variables
during second-stage estimation. Model 2 can be considered an additional
robustness check, which uses the store-level demographics in the first
stage. For Models 1 and 2, we estimated the Probit model for a benchmark,
and for setting an initial value for the nonlinear optimizer to estimate
the Klein-Spady model. The coefficient for bottle size was normalized
to 1. We find that coefficient estimates from Probit estimation and
Klein-Spady estimation are considerably different. We also plot the
estimated conditional density of $\eta_{j,t}$ given $\mathbf{w}_{j,t}$
from each model in Figure \ref{fig:Estimated-Distribution-of}. The
estimated density of $\eta_{j,t}$ given $\mathbf{w}_{j,t}$ is not
even unimodal, which is strong evidence that the unobservable product
characteristic, $\eta_{j,t}$, does not follow a Gaussian distribution.

The primary estimation result is shown in Table \ref{tab:Parameter-Estimates}.
In the first two rows, we present the mean of the implied own price
elasticities from the Marshallian CES and logit demand systems, respectively.
In the logit demand models, coefficients of the log of prices were
positive, and economically and statistically significant, even after
instrumenting for prices using supplier side cost information. As
a result, the own price elasticities are positive, meaning that an
upward-sloping demand curve is estimated. In contrast, the log-linear
estimation of our model with Klein-Spady first-stage estimator returned
the expected signs and magnitudes for coefficients of the log of prices,
and thus, the own price elasticities are negative and the estimated
demand curves are downward-sloping. Estimators assuming a standard
Gaussian distribution on unobservables performed well, despite estimated
distributions of unobservables being far from Gaussian. We argue that
such good performance of models assuming normality is due to the fact
that estimated propensity scores from the Klein-Spady and Probit models
correlate highly, with a correlation coefficient of about 0.7 for
Models 1 and 2. Although this pattern was consistent in all robustness
checks (Appendix \ref{sec:Robustness-Checks}), we are unsure whether
it can be generalized to a different dataset or market.\footnote{\citet{Gandhi2013} also uses bath tissue data from Dominick's database,
using a time series variation of a single chain store. Their estimated
demand function is much more elastic than ours. For example, price
coefficient estimates from simply dropping samples with zero market
shares, which should be biased upward, remain negative. However, the
implication they draw \textendash{} that samples with zero observed
market shares should not be simply dropped \textendash{} is similar
to ours.}

Results provide strong evidence of a consideration set selection process
that has been ignored in demand estimation literature. Ignoring the
consideration set selection process of consumers biases the estimates,
even resulting in an upward-sloping demand curve. Recall the estimation
equation (\ref{eq:loglineardemand}) under the exponential quality
kernel: 
\[
E\left[\ln\left(\frac{\pi_{j,t}}{\pi_{0,t}}\right)|\mathbf{z}_{j,t},\mathbf{w}_{j,t},d_{j,t}=1\right]=-\sigma\phi_{j,t}+\mathbf{x}_{j,t}'\bm{\beta}+E\left[\xi_{j,t}|\mathbf{z}_{j,t},\mathbf{w}_{j,t},d_{j,t}=1\right].
\]
Except for term $E\left[\xi_{j,t}|\mathbf{z}_{j,t},\mathbf{w}_{j,t},d_{j,t}=1\right]$,
the estimation equation is the same as that of the logit demand model
when we dropped samples with zero observed market shares. Columns
(1) (Our Model, K/S, Model 1) and (7) (Logit Model, Drop 0) should
coincide exactly when term $E\left[\xi_{j,t}|\mathbf{z}_{j,t},\mathbf{w}_{j,t},d_{j,t}=1\right]$
is zero, yet this was not the case. $E\left[\xi_{j,t}|\mathbf{z}_{j,t},\mathbf{w}_{j,t},d_{j,t}=1\right]$
is likely positive in our case because consumers select unobservables
$\eta_{j,t}$ and observables $\mathbf{w}_{j,t}$, and $\eta_{j,t}$
correlates highly with $\xi_{j,t}$. Even after instrumenting for
prices, price coefficient estimates are likely to be biased upward
when samples with zero observed market shares are simply dropped.
Imputing small numbers on zero observed market shares might cause
a more serious problem \textendash{} the direction of the bias cannot
be predicted. In contrast to Table \ref{tab:Estimation-Result-of}
in the previous section, Table \ref{tab:Parameter-Estimates} shows
that imputing zero observed market shares with small positive numbers
causes upward bias in price coefficient estimates. We cannot explain
the direction of the bias when zeros are imputed. In Online Appendix
\ref{sec:Robustness-Checks}, we present estimation results for cola
data from different weeks, and for laundry detergent data, with all
results demonstrating the same pattern as that in Table \ref{tab:Parameter-Estimates},
suggesting our findings are robust.

\section{\label{sec:Conclusion}Conclusion}

We develop a semiparametric demand estimation framework based on the
Marshallian demand function derived from the budget-constrained CES
utility maximization problem. Our framework is sufficiently flexible
to incorporate observed and unobserved product characteristics, and
is compatible with the widely used homogeneous and random coefficient
logit models of demand. The framework accommodates zero predicted
and observed market shares with a reasonable microfoundation by separating
intensive and extensive margins, and embedding both margins in a quality
kernel. We account for selection of a consumer's consideration set,
which is unrecognized in the literature. If the consideration set
selection stage is ignored, estimates of price coefficients can be
misleading not only regarding their magnitudes, but also their signs.
We demonstrate that ignoring consideration set selection can even
result in upward-sloping demand curves. A direct extension of our
study is a random coefficient demand estimation framework that can
accommodate zero predicted and observed market shares. When a representative
agent is assumed, the own and cross price elasticities derived from
our model exhibited unrealistic substitution patterns, as in the homogeneous
logit demand model of \citet{Berry1994}. Overcoming such unrealistic
substitution patterns was one of the most important motivations for
development of a random coefficient logit model of demand by \citet{Berry1995}.
Although we provide the microfoundation for a random coefficient CES
demand estimation framework, we do not develop identification and
estimation of model parameters with random coefficients that can accommodate
zero market shares. We leave that extension to future research.%

\newpage{}

\bibliographystyle{econometrica}
\bibliography{Semiparametric_CES_demand_estimation}

\newpage{}

\appendix

\section{\label{sec:Derivation-of-the}Derivation of the Logit Demand System}

We illustrate the derivation of a homogeneous and random coefficient
logit demand systems for completeness. The illustration in this section
largely follows the original presentation of \citet{Berry1994,Berry1995}. 

Let $j\in\mathcal{J}_{t}$, where $\mathcal{J}_{t}$ is a finite set
of alternatives that must contain the numeraire. Individual $i$ in
market $t$ solves the following discrete choice utility maximization
problem: 
\[
\max_{j\in\mathcal{J}_{t}}\left\{ u_{i,j,t}\right\} ,
\]
where the (indirect) utility of choosing alternative $j$ in market
$t$ is:
\begin{equation}
u_{i,j,t}=\alpha_{i}\left(y_{i}-p_{j,t}\right)+\mathbf{x}_{j,t}'\bm{\beta}_{i}+\xi_{j,t}+\epsilon_{i,j,t}.\label{eq:linearutility}
\end{equation}
$\epsilon_{i,j,t}$ follows the i.i.d. Type-I extreme value distribution.
Note that it is also legitimate to specify the utility as:
\[
u_{i,j,t}=-\alpha_{i}\ln p_{j,t}+\mathbf{x}_{j,t}'\bm{\beta}_{i}+\xi_{j,t}+\epsilon_{i,j,t},
\]
given that we stick to the direct utility interpretation of $u_{i,j,t}$
as individual $i$ choosing alternative $j$ in market $t$. The logarithm
can be regarded as a scale adjustment on the level of disutility from
prices.

The coefficients $\left(\alpha_{i},\bm{\beta}_{i}\right)$ might vary
over individuals, and are specified as:
\begin{align*}
\alpha_{i} & :=\alpha+\Pi_{\alpha}\mathbf{q}_{i}+\Sigma_{\alpha}v_{\alpha,i}\\
\bm{\beta}_{i} & :=\bm{\beta}+\bm{\Pi}_{\beta}\mathbf{q}_{i}+\bm{\Sigma}_{\beta}\mathbf{v}_{\beta,i},
\end{align*}
where $\mathbf{q}_{i}$ is the demographic variable, $\mathbf{v}_{i}$
is the vector of a unit normal shock, $\left(\Pi_{\alpha},\bm{\Pi}_{\beta}\right)$
is the correlation component between demographic variables and the
corresponding coefficients, and $\left(\Sigma_{\alpha},\bm{\Sigma}_{\beta}\right)$
represents the covariance structure of the shocks on the coefficients.
The linear utility specification (\ref{eq:linearutility}) becomes:
\begin{eqnarray*}
u_{i,j,t} & = & \alpha_{i}\left(y_{i}-p_{j,t}\right)+\mathbf{x}_{j,t}'\bm{\beta}_{i}+\xi_{j,t}+\epsilon_{i,j,t}\\
 & = & \alpha_{i}y_{i}-\left(\alpha+\Pi_{\alpha}\mathbf{q}_{i}+\Sigma_{\alpha}v_{\alpha,i}\right)p_{j,t}+\mathbf{x}_{j,t}\left(\bm{\beta}+\bm{\Pi}_{\beta}\mathbf{q}_{i}+\bm{\Sigma}_{\beta}\mathbf{v}_{\beta,i}\right)+\xi_{j,t}+\epsilon_{i,j,t}\\
 & = & \alpha_{i}y_{i}+\left(-\alpha p_{j,t}+\mathbf{x}_{j,t}'\bm{\beta}+\xi_{j,t}\right)-\left(\Pi_{\alpha}\mathbf{q}_{i}+\Sigma_{\alpha}v_{\alpha,i}\right)p_{j,t}+\mathbf{x}_{j,t}'\left(\bm{\Pi}_{\beta}\mathbf{q}_{i}+\bm{\Sigma}_{\beta}\mathbf{v}_{\beta,i}\right)+\epsilon_{i,j,t}\\
 & = & \alpha_{i}y_{i}+\left(-\alpha p_{j,t}+\mathbf{x}_{j,t}'\bm{\beta}+\xi_{j,t}\right)+\begin{pmatrix}-p_{j,t} & \mathbf{x}_{j,t}'\end{pmatrix}\left(\bm{\Pi}\mathbf{q}_{i}+\bm{\Sigma}\mathbf{v}_{i}\right)+\epsilon_{i,j,t}\\
 & =: & \alpha_{i}y_{i}+\delta_{j,t}+\mu_{i,j,t}+\epsilon_{i,j,t},
\end{eqnarray*}
where $\delta_{j,t}$ is the mean utility of alternative $j$ that
is common to every individual in market $t$, and $\mu_{i,j,t}$ is
the individual specific structural utility component. For the log-linear
specification, one can simply replace the term $p_{j,t}$ with $\ln p_{j,t}$.

Given the assumption that $\epsilon_{i,j,t}$ follows i.i.d. Type-I
extreme value distribution, the individual choice probability $\Pr\left(i\rightarrow j|t\right)$
becomes:

\[
\Pr\left(i\rightarrow j|t\right)=\frac{\exp\left(\delta_{j,t}+\mu_{i,j,t}\right)}{\sum_{k\in\mathcal{J}_{t}}\exp\left(\delta_{k,t}+\mu_{i,k,t}\right)}.
\]
This individual choice probability is taken as the individual predicted
quantity share $\pi_{i,j,t}$. Given distributions of the demographics
$F\left(\mathbf{z}_{i}\right)$ and of shocks on the preference parameter
$F\left(\mathbf{v}_{i}\right)$, the predicted quantity market share
of good $j$ is aggregated as:
\begin{eqnarray}
\pi_{j,t} & = & \int\int\pi_{i,j,t}dF\left(\mathbf{z}_{i}\right)dF\left(\mathbf{v}_{i}\right)\label{eq:zxcdjfklsdaf}\\
 & = & \int\int\frac{\exp\left(\delta_{j,t}+\mu_{i,j,t}\right)}{\sum_{k\in\mathcal{J}_{t}}\exp\left(\delta_{k,t}+\mu_{i,k,t}\right)}dF\left(\mathbf{z}_{i}\right)dF\left(\mathbf{v}_{i}\right).\nonumber 
\end{eqnarray}
If $\alpha_{i}=\alpha$ and $\bm{\beta}_{i}=\bm{\beta}$, which implies
that the preference is homogeneous across individuals, the model reduces
to the homogeneous logit demand model.

By definition, the predicted market share $\pi_{j,t}$ is:
\[
\pi_{j,t}:=\frac{q_{j,t}}{\sum_{k\in\mathcal{J}_{t}}q_{k,t}}.
\]
This system of predicted quantity market shares for $\#\left(\mathcal{J}_{t}\right)$
alternatives in a market $t$ provides only $\#\left(\mathcal{J}_{t}\right)-1$
restrictions on the system of quantity demand $\mathbf{q}_{t}$. An
additional restriction is required, and \citet{Berry1994,Berry1995}
impose a fixed market size assumption to derive the quantity demand;
denominator $\sum_{k\in\mathcal{J}_{t}}q_{k,t}$ is regarded as fixed
at some level $M$.

\section{\label{sec:Implementation-Details}Implementation Details}

We use the Gaussian kernel during first- and second-stage estimation.
For tractability, higher-order kernels are not used. The bandwidth,
$h_{n}$, for the Klein-Spady estimator is $h_{n}=\mbox{std}\left(\mathbf{w}_{j}'\hat{\bm{\delta}}_{\mbox{Probit}}\right)C_{1}n^{-\frac{1}{7}}$.
The rate $n^{-\frac{1}{7}}$ follows the original suggestion from
\citet{Klein1993}. Bandwidth for the second-stage \citet{Powell2001}
estimator is $h_{n}=\mbox{std}\left(\mathbf{w}_{j}'\hat{\bm{\delta}}_{\mbox{KS}}\right)C_{2}n^{-\frac{1}{7}}$,
where $\hat{\bm{\delta}}_{\mbox{KS}}$ is the Klein-Spady estimator
from the first stage. We use the tuning parameter $C_{1}=C_{2}=1$
in Section \ref{sec:Monte-Carlo-Simulations} and $C_{1}=C_{2}=0.5$
in Section \ref{sec:Empirical-Example}. 

We tried 100 randomly generated starting values in the first stage
Klein-Spady estimation to guard against the argument that the optimization
routine stopped at the local minima. The randomly generated initial
values follow the distribution $\mathcal{N}\left(\hat{\bm{\delta}}_{\mbox{Probit}},\frac{1}{5}\mbox{diag}\left(\sqrt{\left|\hat{\bm{\delta}}_{\mbox{Probit}}\right|}\right)\right)$.
We also tried several tuning parameters to assess robustness. First-stage
parameter estimates varied considerably regarding the choice of tuning
parameters and bandwidth, whereas second-stage parameter estimates,
which are our primary interest, were robust to the choice of bandwidth
and initial values for nonlinear optimization. 

For the simple Heckman estimator with endogeneity, we computed standard
errors that account for the fact that the inverse Mills ratio is a
generated regressor. Details on the covariance formula of the estimator
can be found in \citet{Newey1994}. Because finding the inverse Mills
ratio is fast, one can also consider the bootstrapped standard errors
for the Heckman estimator. Finally, we tried both IPOPT and KNITRO,
which are state-of-the-art, derivatives-based, nonlinear optimizers
for nonlinear optimization. Results were robust to choice of optimizer. 

\newpage{}

\section{\label{sec:Robustness-Checks}For Online Publication: Robustness
Checks}

\subsection{\label{subsec:Cola-Data-for}Cola Demand for Week 382}

We estimate the same models as in Section \ref{sec:Empirical-Example}
using cola data from a different week. We use data from week 382 (January
1 through 9, 1997). In Tables \ref{tab:First-stage-Parameter-Estimates-1}
and \ref{tab:Parameter-Estimates-1}, we repeat the estimation procedure
from Tables \ref{tab:First-stage-Parameter-Estimates} and \ref{tab:Parameter-Estimates}. 

\begin{table}[!tbph]
\caption{\label{tab:First-stage-Parameter-Estimates-1}First-stage Parameter
Estimates $\hat{\bm{\delta}}$ }

\begin{centering}
\begin{tabular}{cccccc}
\hline 
 & \multicolumn{2}{c}{{\footnotesize{}Model 1}} &  & \multicolumn{2}{c}{{\footnotesize{}Model 2}}\tabularnewline
\cline{2-3} \cline{5-6} 
 & {\footnotesize{}Probit}  & {\footnotesize{}Klein-Spady}  &  & {\footnotesize{}Probit}  & {\footnotesize{}Klein-Spady}\tabularnewline
{\footnotesize{}$\mathbf{w}_{j,t}$}  & {\footnotesize{}(1)} & {\footnotesize{}(2)} &  & {\footnotesize{}(3)} & {\footnotesize{}(4)}\tabularnewline
\hline 
{\footnotesize{}Bottle Size}  & {\footnotesize{}$\begin{array}{c}
1\\
\left(0.188\right)
\end{array}$}  & {\footnotesize{}$\begin{array}{c}
1\\
\left(-\right)
\end{array}$} &  & {\footnotesize{}$\begin{array}{c}
1\\
\left(0.193\right)
\end{array}$}  & {\footnotesize{}$\begin{array}{c}
1\\
\left(-\right)
\end{array}$} \tabularnewline
{\footnotesize{}\# Bottles per Bundle}  & {\footnotesize{}$\begin{array}{c}
-5.867\\
\left(0.429\right)
\end{array}$}  & {\footnotesize{}$\begin{array}{c}
-46.247\\
\left(12.736\right)
\end{array}$} &  & {\footnotesize{}$\begin{array}{c}
-5.825\\
\left(0.426\right)
\end{array}$}  & {\footnotesize{}$\begin{array}{c}
-157.181\\
\left(23.221\right)
\end{array}$} \tabularnewline
{\footnotesize{}Diet}  & {\footnotesize{}$\begin{array}{c}
14.892\\
\left(5.016\right)
\end{array}$}  & {\footnotesize{}$\begin{array}{c}
7.834\\
\left(3.170\right)
\end{array}$} &  & {\footnotesize{}$\begin{array}{c}
14.674\\
\left(4.986\right)
\end{array}$}  & {\footnotesize{}$\begin{array}{c}
5.802\\
\left(5.447\right)
\end{array}$} \tabularnewline
{\footnotesize{}Caffeine Free}  & {\footnotesize{}$\begin{array}{c}
-22.949\\
\left(5.971\right)
\end{array}$}  & {\footnotesize{}$\begin{array}{c}
43.032\\
\left(4.613\right)
\end{array}$} &  & {\footnotesize{}$\begin{array}{c}
-22.501\\
\left(5.938\right)
\end{array}$}  & {\footnotesize{}$\begin{array}{c}
-335.901\\
\left(44.885\right)
\end{array}$} \tabularnewline
{\footnotesize{}Cherry}  & {\footnotesize{}$\begin{array}{c}
-68.882\\
\left(11.761\right)
\end{array}$}  & {\footnotesize{}$\begin{array}{c}
-231.403\\
\left(71.908\right)
\end{array}$} &  & {\footnotesize{}$\begin{array}{c}
-69.030\\
\left(11.635\right)
\end{array}$}  & {\footnotesize{}$\begin{array}{c}
-188.603\\
\left(23.457\right)
\end{array}$} \tabularnewline
{\footnotesize{}Coke}  & {\footnotesize{}$\begin{array}{c}
70.670\\
\left(9.636\right)
\end{array}$}  & {\footnotesize{}$\begin{array}{c}
122.727\\
\left(17.578\right)
\end{array}$} &  & {\footnotesize{}$\begin{array}{c}
70.572\\
\left(9.577\right)
\end{array}$}  & {\footnotesize{}$\begin{array}{c}
262.402\\
\left(33.91\right)
\end{array}$} \tabularnewline
{\footnotesize{}Pepsi}  & {\footnotesize{}$\begin{array}{c}
48.249\\
\left(5.888\right)
\end{array}$}  & {\footnotesize{}$\begin{array}{c}
-150.090\\
\left(56.156\right)
\end{array}$} &  & {\footnotesize{}$\begin{array}{c}
47.706\\
\left(5.839\right)
\end{array}$}  & {\footnotesize{}$\begin{array}{c}
377.354\\
\left(45.846\right)
\end{array}$} \tabularnewline
{\footnotesize{}Promo}  & {\footnotesize{}$\begin{array}{c}
126.413\\
\left(7.699\right)
\end{array}$}  & {\footnotesize{}$\begin{array}{c}
866.065\\
\left(229.737\right)
\end{array}$} &  & {\footnotesize{}$\begin{array}{c}
124.981\\
\left(7.633\right)
\end{array}$}  & {\footnotesize{}$\begin{array}{c}
1007.944\\
\left(139.527\right)
\end{array}$} \tabularnewline
{\footnotesize{}\% Blacks and Hispanics}  & {\footnotesize{}-}  & {\footnotesize{}-} &  & {\footnotesize{}$\begin{array}{c}
24.100\\
\left(22.881\right)
\end{array}$}  & {\footnotesize{}$\begin{array}{c}
30.682\\
\left(23.885\right)
\end{array}$} \tabularnewline
{\footnotesize{}\% College Graduates}  & {\footnotesize{}-}  & {\footnotesize{}-} &  & {\footnotesize{}$\begin{array}{c}
-45.023\\
\left(22.069\right)
\end{array}$}  & {\footnotesize{}$\begin{array}{c}
-66.658\\
\left(21.641\right)
\end{array}$} \tabularnewline
{\footnotesize{}Log Median Income}  & {\footnotesize{}-}  & {\footnotesize{}-} &  & {\footnotesize{}$\begin{array}{c}
-284.144\\
\left(100.267\right)
\end{array}$}  & {\footnotesize{}$\begin{array}{c}
-226.55\\
\left(100.23\right)
\end{array}$} \tabularnewline
 &  &  &  &  & \tabularnewline
{\footnotesize{}$D$}  & {\footnotesize{}$3226$}  & {\footnotesize{}$3226$}  &  & {\footnotesize{}$3226$}  & {\footnotesize{}$3226$}\tabularnewline
{\footnotesize{}$N$}  & {\footnotesize{}$4337$}  & {\footnotesize{}$4337$}  &  & {\footnotesize{}$4337$}  & {\footnotesize{}$4337$}\tabularnewline
\hline 
\end{tabular}\medskip{}
\par\end{centering}
{\footnotesize{}Note. (i) $D$ is the number of non-zero market share
observations, and $N$ is the sample size. (ii) The asymptotic standard
error estimates appear in parentheses. (iii) The unit of bottle size
is liquid ounces. (iv) We normalized the coefficients of the Bottle
Size variable to one. }{\footnotesize\par}
\end{table}

\begin{sidewaystable}[H]
\caption{\label{tab:Parameter-Estimates-1}Second-stage Parameter Estimates
$\left(\hat{\sigma},\hat{\bm{\beta}}\right)$}

\begin{centering}
\begin{tabular}{ccccccccccccc}
\hline 
 & \multicolumn{2}{c}{{\footnotesize{}Our Model, K/S }} &  & \multicolumn{2}{c}{{\footnotesize{}Our Model, Probit}} &  & \multicolumn{2}{c}{{\footnotesize{}Heckman Correction}} &  & \multicolumn{3}{c}{{\footnotesize{}Logit Model}}\tabularnewline
\cline{2-3} \cline{5-6} \cline{8-9} \cline{11-13} 
 & {\footnotesize{}Model 1}  & {\footnotesize{}Model 2}  &  & {\footnotesize{}Model 1}  & {\footnotesize{}Model 2}  &  & {\footnotesize{}Model 1}  & {\footnotesize{}Model 2}  &  & {\footnotesize{}Drop 0}  & {\footnotesize{}$10^{-8}$}  & {\footnotesize{}$10^{-4}$}\tabularnewline
{\footnotesize{}$\left(\phi_{j,t},\mathbf{x}_{j,t}\right)$}  & {\footnotesize{}(1)} & {\footnotesize{}(2)} &  & {\footnotesize{}(3)} & {\footnotesize{}(4)} &  & {\footnotesize{}(5)} & {\footnotesize{}(6)} &  & {\footnotesize{}(7)} & {\footnotesize{}(8)} & {\footnotesize{}(9)}\tabularnewline
\hline 
{\footnotesize{}Log Price ($-\sigma$)}  & {\footnotesize{}$\begin{array}{c}
-1.175\\
\left(0.039\right)
\end{array}$}  & {\footnotesize{}$\begin{array}{c}
-1.254\\
\left(0.035\right)
\end{array}$}  &  & {\footnotesize{}$\begin{array}{c}
-1.266\\
\left(0.045\right)
\end{array}$}  & {\footnotesize{}$\begin{array}{c}
-1.318\\
\left(0.038\right)
\end{array}$}  &  & {\footnotesize{}$\begin{array}{c}
-1.272\\
\left(0.041\right)
\end{array}$}  & {\footnotesize{}$\begin{array}{c}
-1.339\\
\left(0.038\right)
\end{array}$}  &  & {\footnotesize{}$\begin{array}{c}
4.367\\
\left(0.132\right)
\end{array}$} & {\footnotesize{}$\begin{array}{c}
5.348\\
\left(0.168\right)
\end{array}$} & {\footnotesize{}$\begin{array}{c}
4.055\\
\left(0.085\right)
\end{array}$}\tabularnewline
{\footnotesize{}Bottle Size}  & {\footnotesize{}$\begin{array}{c}
0.010\\
\left(0.004\right)
\end{array}$}  & {\footnotesize{}$\begin{array}{c}
0.009\\
\left(0.002\right)
\end{array}$}  &  & {\footnotesize{}$\begin{array}{c}
-0.003\\
\left(0.006\right)
\end{array}$}  & {\footnotesize{}$\begin{array}{c}
-0.002\\
\left(0.003\right)
\end{array}$}  &  & {\footnotesize{}$\begin{array}{c}
-0.002\\
\left(0.001\right)
\end{array}$}  & {\footnotesize{}$\begin{array}{c}
0.002\\
\left(0.001\right)
\end{array}$}  &  & {\footnotesize{}$\begin{array}{c}
0.107\\
\left(0.005\right)
\end{array}$} & {\footnotesize{}$\begin{array}{c}
0.136\\
\left(0.007\right)
\end{array}$} & {\footnotesize{}$\begin{array}{c}
0.095\\
\left(0.004\right)
\end{array}$}\tabularnewline
{\footnotesize{}\# Bottles per Bundle}  & {\footnotesize{}$\begin{array}{c}
0.053\\
\left(0.003\right)
\end{array}$}  & {\footnotesize{}$\begin{array}{c}
-0.001\\
\left(0.009\right)
\end{array}$}  &  & {\footnotesize{}$\begin{array}{c}
0.046\\
\left(0.004\right)
\end{array}$}  & {\footnotesize{}$\begin{array}{c}
0.011\\
\left(0.014\right)
\end{array}$}  &  & {\footnotesize{}$\begin{array}{c}
0.036\\
\left(0.003\right)
\end{array}$}  & {\footnotesize{}$\begin{array}{c}
-0.018\\
\left(0.004\right)
\end{array}$}  &  & {\footnotesize{}$\begin{array}{c}
0.246\\
\left(0.015\right)
\end{array}$} & {\footnotesize{}$\begin{array}{c}
0.192\\
\left(0.019\right)
\end{array}$} & {\footnotesize{}$\begin{array}{c}
0.193\\
\left(0.010\right)
\end{array}$}\tabularnewline
{\footnotesize{}Diet}  & {\footnotesize{}$\begin{array}{c}
-0.114\\
\left(0.036\right)
\end{array}$}  & {\footnotesize{}$\begin{array}{c}
-0.127\\
\left(0.032\right)
\end{array}$}  &  & {\footnotesize{}$\begin{array}{c}
-0.113\\
\left(0.087\right)
\end{array}$}  & {\footnotesize{}$\begin{array}{c}
-0.119\\
\left(0.042\right)
\end{array}$}  &  & {\footnotesize{}$\begin{array}{c}
-0.275\\
\left(0.048\right)
\end{array}$}  & {\footnotesize{}$\begin{array}{c}
-0.155\\
\left(0.035\right)
\end{array}$}  &  & {\footnotesize{}$\begin{array}{c}
1.132\\
\left(0.136\right)
\end{array}$} & {\footnotesize{}$\begin{array}{c}
1.360\\
\left(0.197\right)
\end{array}$} & {\footnotesize{}$\begin{array}{c}
0.891\\
\left(0.099\right)
\end{array}$}\tabularnewline
{\footnotesize{}Caffeine Free}  & {\footnotesize{}$\begin{array}{c}
-1.108\\
\left(0.053\right)
\end{array}$}  & {\footnotesize{}$\begin{array}{c}
-1.215\\
\left(0.041\right)
\end{array}$}  &  & {\footnotesize{}$\begin{array}{c}
-1.148\\
\left(0.114\right)
\end{array}$}  & {\footnotesize{}$\begin{array}{c}
-1.111\\
\left(0.054\right)
\end{array}$}  &  & {\footnotesize{}$\begin{array}{c}
-0.851\\
\left(0.053\right)
\end{array}$}  & {\footnotesize{}$\begin{array}{c}
-1.146\\
\left(0.040\right)
\end{array}$}  &  & {\footnotesize{}$\begin{array}{c}
-1.699\\
\left(0.135\right)
\end{array}$} & {\footnotesize{}$\begin{array}{c}
-2.740\\
\left(0.204\right)
\end{array}$} & {\footnotesize{}$\begin{array}{c}
-1.605\\
\left(0.103\right)
\end{array}$}\tabularnewline
{\footnotesize{}Cherry}  & {\footnotesize{}$\begin{array}{c}
-2.222\\
\left(0.272\right)
\end{array}$}  & {\footnotesize{}$\begin{array}{c}
-1.889\\
\left(0.105\right)
\end{array}$}  &  & {\footnotesize{}$\begin{array}{c}
-1.772\\
\left(0.324\right)
\end{array}$}  & {\footnotesize{}$\begin{array}{c}
-1.749\\
\left(0.132\right)
\end{array}$}  &  & {\footnotesize{}$\begin{array}{c}
-0.725\\
\left(0.218\right)
\end{array}$}  & {\footnotesize{}$\begin{array}{c}
-1.696\\
\left(0.120\right)
\end{array}$}  &  & {\footnotesize{}$\begin{array}{c}
-3.153\\
\left(0.410\right)
\end{array}$} & {\footnotesize{}$\begin{array}{c}
-6.635\\
\left(0.485\right)
\end{array}$} & {\footnotesize{}$\begin{array}{c}
-3.031\\
\left(0.245\right)
\end{array}$}\tabularnewline
{\footnotesize{}Coke}  & {\footnotesize{}$\begin{array}{c}
1.670\\
\left(0.136\right)
\end{array}$}  & {\footnotesize{}$\begin{array}{c}
1.461\\
\left(0.113\right)
\end{array}$}  &  & {\footnotesize{}$\begin{array}{c}
1.415\\
\left(0.392\right)
\end{array}$}  & {\footnotesize{}$\begin{array}{c}
1.271\\
\left(0.135\right)
\end{array}$}  &  & {\footnotesize{}$\begin{array}{c}
0.445\\
\left(0.123\right)
\end{array}$}  & {\footnotesize{}$\begin{array}{c}
1.178\\
\left(0.098\right)
\end{array}$}  &  & {\footnotesize{}$\begin{array}{c}
3.270\\
\left(0.259\right)
\end{array}$} & {\footnotesize{}$\begin{array}{c}
5.801\\
\left(0.361\right)
\end{array}$} & {\footnotesize{}$\begin{array}{c}
2.888\\
\left(0.182\right)
\end{array}$}\tabularnewline
{\footnotesize{}Pepsi}  & {\footnotesize{}$\begin{array}{c}
0.844\\
\left(0.126\right)
\end{array}$}  & {\footnotesize{}$\begin{array}{c}
0.942\\
\left(0.062\right)
\end{array}$}  &  & {\footnotesize{}$\begin{array}{c}
1.045\\
\left(0.237\right)
\end{array}$}  & {\footnotesize{}$\begin{array}{c}
0.917\\
\left(0.079\right)
\end{array}$}  &  & {\footnotesize{}$\begin{array}{c}
0.289\\
\left(0.069\right)
\end{array}$}  & {\footnotesize{}$\begin{array}{c}
0.798\\
\left(0.052\right)
\end{array}$}  &  & {\footnotesize{}$\begin{array}{c}
3.476\\
\left(0.202\right)
\end{array}$} & {\footnotesize{}$\begin{array}{c}
5.612\\
\left(0.264\right)
\end{array}$} & {\footnotesize{}$\begin{array}{c}
3.030\\
\left(0.134\right)
\end{array}$}\tabularnewline
{\footnotesize{}Promo}  & {\footnotesize{}-}  & {\footnotesize{}$\begin{array}{c}
1.510\\
\left(0.063\right)
\end{array}$}  &  & {\footnotesize{}-}  & {\footnotesize{}$\begin{array}{c}
0.645\\
\left(0.257\right)
\end{array}$}  &  & {\footnotesize{}-}  & {\footnotesize{}$\begin{array}{c}
1.256\\
\left(0.066\right)
\end{array}$}  &  & {\footnotesize{}-} & {\footnotesize{}-} & {\footnotesize{}-}\tabularnewline
 &  &  &  &  &  &  &  &  &  &  &  & \tabularnewline
{\footnotesize{}$D$}  & {\footnotesize{}$3226$}  & {\footnotesize{}$3226$}  &  & {\footnotesize{}$3226$}  & {\footnotesize{}$3226$}  &  & {\footnotesize{}$3226$}  & {\footnotesize{}$3226$}  &  & {\footnotesize{}$3226$}  & {\footnotesize{}$4118$}  & {\footnotesize{}$4118$}\tabularnewline
{\footnotesize{}$N$}  & {\footnotesize{}$4337$}  & {\footnotesize{}$4337$}  &  & {\footnotesize{}$4337$}  & {\footnotesize{}$4337$}  &  & {\footnotesize{}$4337$}  & {\footnotesize{}$4337$}  &  & {\footnotesize{}$3226$}  & {\footnotesize{}$4118$}  & {\footnotesize{}$4118$}\tabularnewline
\hline 
\end{tabular}\medskip{}
\par\end{centering}
{\footnotesize{}Note. (i) For columns Our Model, K/S and Our Model,
Probit, results from the Klein-Spady and Probit estimators in Table
\ref{tab:First-stage-Parameter-Estimates-1} were used for the first-stage
estimator, respectively. A pairwise differenced weighted instrumental
variable estimator was then used during the second stage. For the
Heckman Correction column, the Probit was used during the first stage,
and the Heckman's selection correction estimator with the inverse
Mills ratio as an additional regressor was used during the second
stage. (ii) $D$ is the number of non-zero market share observations,
and $N$ is the effective sample size. (iii) Asymptotic standard error
estimates are in parentheses. (iv) The unit of bottle size is liquid
ounces. (v) Because Dominick's did not record the price and cost when
sales were zero, when estimating the $10^{-8}$ and $10^{-4}$ columns,
we used average prices and costs of the same product with the same
promotion statuses from other stores.}{\footnotesize\par}
\end{sidewaystable}

\newpage{}

\subsection{\label{subsec:Cola-Demand-for}Cola Demand for Week 278}

We repeat estimation of cola data from week 278 (January 5 through
11, 1995). All results demonstrated the same pattern as in previous
sections. Although not tabulated here, we also examined data from
many other weeks, and results were robust.

\begin{table}[!tbph]
\caption{\label{tab:First-stage-Parameter-Estimates-2}First-stage Parameter
Estimates $\hat{\bm{\delta}}$ }

\begin{centering}
\begin{tabular}{cccccc}
\hline 
 & \multicolumn{2}{c}{{\footnotesize{}Model 1}} &  & \multicolumn{2}{c}{{\footnotesize{}Model 2}}\tabularnewline
\cline{2-6} 
 & {\footnotesize{}Probit}  & {\footnotesize{}Klein-Spady}  &  & {\footnotesize{}Probit}  & {\footnotesize{}Klein-Spady}\tabularnewline
{\footnotesize{}$\mathbf{w}_{j,t}$}  & {\footnotesize{}(1)} & {\footnotesize{}(2)} &  & {\footnotesize{}(3)} & {\footnotesize{}(4)}\tabularnewline
\hline 
{\footnotesize{}Bottle Size}  & {\footnotesize{}$\begin{array}{c}
1\\
\left(0.170\right)
\end{array}$}  & {\footnotesize{}$\begin{array}{c}
1\\
\left(-\right)
\end{array}$} &  & {\footnotesize{}$\begin{array}{c}
1\\
\left(0.169\right)
\end{array}$}  & {\footnotesize{}$\begin{array}{c}
1\\
\left(-\right)
\end{array}$} \tabularnewline
{\footnotesize{}\# Bottles per Bundle}  & {\footnotesize{}$\begin{array}{c}
-5.334\\
\left(0.489\right)
\end{array}$}  & {\footnotesize{}$\begin{array}{c}
-166.569\\
\left(140.200\right)
\end{array}$} &  & {\footnotesize{}$\begin{array}{c}
-5.289\\
\left(0.487\right)
\end{array}$}  & {\footnotesize{}$\begin{array}{c}
-68.799\\
\left(5.218\right)
\end{array}$} \tabularnewline
{\footnotesize{}Diet}  & {\footnotesize{}$\begin{array}{c}
-76.147\\
\left(6.761\right)
\end{array}$}  & {\footnotesize{}$\begin{array}{c}
-1520.618\\
\left(1289.560\right)
\end{array}$} &  & {\footnotesize{}$\begin{array}{c}
-75.673\\
\left(6.774\right)
\end{array}$}  & {\footnotesize{}$\begin{array}{c}
-207.218\\
\left(15.787\right)
\end{array}$} \tabularnewline
{\footnotesize{}Caffeine Free}  & {\footnotesize{}$\begin{array}{c}
74.187\\
\left(8.674\right)
\end{array}$}  & {\footnotesize{}$\begin{array}{c}
1565.247\\
\left(1292.587\right)
\end{array}$} &  & {\footnotesize{}$\begin{array}{c}
75.038\\
\left(8.743\right)
\end{array}$}  & {\footnotesize{}$\begin{array}{c}
252.376\\
\left(18.212\right)
\end{array}$} \tabularnewline
{\footnotesize{}Cherry}  & {\footnotesize{}$\begin{array}{c}
-223.710\\
\left(108.580\right)
\end{array}$}  & {\footnotesize{}$\begin{array}{c}
-255.074\\
\left(12145.276\right)
\end{array}$} &  & {\footnotesize{}$\begin{array}{c}
-224.050\\
\left(107.821\right)
\end{array}$}  & {\footnotesize{}$\begin{array}{c}
-1406.267\\
\left(106.027\right)
\end{array}$} \tabularnewline
{\footnotesize{}Coke}  & {\footnotesize{}$\begin{array}{c}
151.949\\
\left(17.472\right)
\end{array}$}  & {\footnotesize{}$\begin{array}{c}
83.965\\
\left(6.953\right)
\end{array}$} &  & {\footnotesize{}$\begin{array}{c}
152.293\\
\left(17.382\right)
\end{array}$}  & {\footnotesize{}$\begin{array}{c}
1460.142\\
\left(106.708\right)
\end{array}$} \tabularnewline
{\footnotesize{}Pepsi}  & {\footnotesize{}$\begin{array}{c}
18.408\\
\left(7.076\right)
\end{array}$}  & {\footnotesize{}$\begin{array}{c}
-751.111\\
\left(643.282\right)
\end{array}$} &  & {\footnotesize{}$\begin{array}{c}
18.103\\
\left(7.046\right)
\end{array}$}  & {\footnotesize{}$\begin{array}{c}
92.893\\
\left(5.724\right)
\end{array}$} \tabularnewline
{\footnotesize{}Promo}  & {\footnotesize{}$\begin{array}{c}
305.656\\
\left(7.716\right)
\end{array}$}  & {\footnotesize{}$\begin{array}{c}
4718.847\\
\left(3869.202\right)
\end{array}$} &  & {\footnotesize{}$\begin{array}{c}
303.695\\
\left(7.697\right)
\end{array}$}  & {\footnotesize{}$\begin{array}{c}
2118.076\\
\left(156.271\right)
\end{array}$} \tabularnewline
{\footnotesize{}\% Blacks and Hispanics}  & {\footnotesize{}-}  & {\footnotesize{}-} &  & {\footnotesize{}$\begin{array}{c}
1.897\\
\left(25.054\right)
\end{array}$}  & {\footnotesize{}$\begin{array}{c}
20.684\\
\left(22.897\right)
\end{array}$} \tabularnewline
{\footnotesize{}\% College Graduates}  & {\footnotesize{}-}  & {\footnotesize{}-} &  & {\footnotesize{}$\begin{array}{c}
-60.347\\
\left(27.667\right)
\end{array}$}  & {\footnotesize{}$\begin{array}{c}
-44.537\\
\left(21.267\right)
\end{array}$} \tabularnewline
{\footnotesize{}Log Median Income}  & {\footnotesize{}-}  & {\footnotesize{}-} &  & {\footnotesize{}$\begin{array}{c}
-213.245\\
\left(112.231\right)
\end{array}$}  & {\footnotesize{}$\begin{array}{c}
-103.183\\
\left(92.389\right)
\end{array}$} \tabularnewline
 &  &  &  &  & \tabularnewline
{\footnotesize{}$D$}  & {\footnotesize{}$3667$}  & {\footnotesize{}$3667$}  &  & {\footnotesize{}$3667$}  & {\footnotesize{}$3667$}\tabularnewline
{\footnotesize{}$N$}  & {\footnotesize{}$5185$}  & {\footnotesize{}$5185$}  &  & {\footnotesize{}$5185$}  & {\footnotesize{}$5185$}\tabularnewline
\hline 
\end{tabular}\medskip{}
\par\end{centering}
{\footnotesize{}Note. (i) $D$ is the number of non-zero market shares
observations, and $N$ is the sample size. (ii) Asymptotic standard
error estimates appear in parentheses. (iii) The unit of bottle size
is liquid ounces. (iv) We normalized the coefficients of the Bottle
Size variable to one. }{\footnotesize\par}
\end{table}

\begin{sidewaystable}[H]
\caption{\label{tab:Second-stage-Parameter-Estimates}Second-stage Parameter
Estimates $\left(\hat{\sigma},\hat{\bm{\beta}}\right)$}

\begin{centering}
\begin{tabular}{ccccccccccccc}
\hline 
 & \multicolumn{2}{c}{{\footnotesize{}Our Model, K/S }} &  & \multicolumn{2}{c}{{\footnotesize{}Our Model, Probit}} &  & \multicolumn{2}{c}{{\footnotesize{}Heckman Correction}} &  & \multicolumn{3}{c}{{\footnotesize{}Logit Model}}\tabularnewline
\cline{2-3} \cline{5-6} \cline{8-9} \cline{11-13} 
 & {\footnotesize{}Model 1}  & {\footnotesize{}Model 2}  &  & {\footnotesize{}Model 1}  & {\footnotesize{}Model 2}  &  & {\footnotesize{}Model 1}  & {\footnotesize{}Model 2}  &  & {\footnotesize{}Drop 0}  & {\footnotesize{}$10^{-8}$}  & {\footnotesize{}$10^{-4}$}\tabularnewline
{\footnotesize{}$\left(\phi_{j,t},\mathbf{x}_{j,t}\right)$} & {\footnotesize{}(1)} & {\footnotesize{}(2)} &  & {\footnotesize{}(3)} & {\footnotesize{}(4)} &  & {\footnotesize{}(5)} & {\footnotesize{}(6)} &  & {\footnotesize{}(7)} & {\footnotesize{}(8)} & {\footnotesize{}(9)}\tabularnewline
\hline 
{\footnotesize{}Log Price ($-\sigma$)}  & {\footnotesize{}$\begin{array}{c}
-0.918\\
\left(0.176\right)
\end{array}$}  & {\footnotesize{}$\begin{array}{c}
-1.225\\
\left(0.068\right)
\end{array}$}  &  & {\footnotesize{}$\begin{array}{c}
-0.261\\
\left(0.110\right)
\end{array}$}  & {\footnotesize{}$\begin{array}{c}
-0.261\\
\left(0.117\right)
\end{array}$}  &  & {\footnotesize{}$\begin{array}{c}
-0.969\\
\left(0.040\right)
\end{array}$}  & {\footnotesize{}$\begin{array}{c}
-0.991\\
\left(0.043\right)
\end{array}$}  &  & {\footnotesize{}$\begin{array}{c}
18.789\\
\left(3.213\right)
\end{array}$} & {\footnotesize{}$\begin{array}{c}
20.461\\
\left(2.736\right)
\end{array}$} & {\footnotesize{}$\begin{array}{c}
16.460\\
\left(2.174\right)
\end{array}$}\tabularnewline
{\footnotesize{}Bottle Size}  & {\footnotesize{}$\begin{array}{c}
0.008\\
\left(0.001\right)
\end{array}$}  & {\footnotesize{}$\begin{array}{c}
0.000\\
\left(0.002\right)
\end{array}$}  &  & {\footnotesize{}$\begin{array}{c}
0.028\\
\left(0.003\right)
\end{array}$}  & {\footnotesize{}$\begin{array}{c}
0.017\\
\left(0.005\right)
\end{array}$}  &  & {\footnotesize{}$\begin{array}{c}
0.013\\
\left(0.001\right)
\end{array}$}  & {\footnotesize{}$\begin{array}{c}
0.012\\
\left(0.001\right)
\end{array}$}  &  & {\footnotesize{}$\begin{array}{c}
0.742\\
\left(0.137\right)
\end{array}$} & {\footnotesize{}$\begin{array}{c}
0.817\\
\left(0.120\right)
\end{array}$} & {\footnotesize{}$\begin{array}{c}
0.654\\
\left(0.095\right)
\end{array}$}\tabularnewline
{\footnotesize{}\# Bottles per Bundle}  & {\footnotesize{}$\begin{array}{c}
0.047\\
\left(0.006\right)
\end{array}$}  & {\footnotesize{}$\begin{array}{c}
-0.018\\
\left(0.019\right)
\end{array}$}  &  & {\footnotesize{}$\begin{array}{c}
0.097\\
\left(0.012\right)
\end{array}$}  & {\footnotesize{}$\begin{array}{c}
0.153\\
\left(0.026\right)
\end{array}$}  &  & {\footnotesize{}$\begin{array}{c}
0.054\\
\left(0.003\right)
\end{array}$}  & {\footnotesize{}$\begin{array}{c}
0.057\\
\left(0.003\right)
\end{array}$}  &  & {\footnotesize{}$\begin{array}{c}
2.169\\
\left(0.401\right)
\end{array}$} & {\footnotesize{}$\begin{array}{c}
2.310\\
\left(0.345\right)
\end{array}$} & {\footnotesize{}$\begin{array}{c}
1.871\\
\left(0.274\right)
\end{array}$}\tabularnewline
{\footnotesize{}Diet}  & {\footnotesize{}$\begin{array}{c}
-0.213\\
\left(0.094\right)
\end{array}$}  & {\footnotesize{}$\begin{array}{c}
0.088\\
\left(0.064\right)
\end{array}$}  &  & {\footnotesize{}$\begin{array}{c}
0.007\\
\left(0.124\right)
\end{array}$}  & {\footnotesize{}$\begin{array}{c}
0.826\\
\left(0.404\right)
\end{array}$}  &  & {\footnotesize{}$\begin{array}{c}
-0.127\\
\left(0.037\right)
\end{array}$}  & {\footnotesize{}$\begin{array}{c}
-0.070\\
\left(0.042\right)
\end{array}$}  &  & {\footnotesize{}$\begin{array}{c}
9.715\\
\left(2.054\right)
\end{array}$} & {\footnotesize{}$\begin{array}{c}
10.045\\
\left(1.502\right)
\end{array}$} & {\footnotesize{}$\begin{array}{c}
7.090\\
\left(1.193\right)
\end{array}$}\tabularnewline
{\footnotesize{}Caffeine Free}  & {\footnotesize{}$\begin{array}{c}
-1.437\\
\left(0.106\right)
\end{array}$}  & {\footnotesize{}$\begin{array}{c}
-1.536\\
\left(0.070\right)
\end{array}$}  &  & {\footnotesize{}$\begin{array}{c}
-1.554\\
\left(0.108\right)
\end{array}$}  & {\footnotesize{}$\begin{array}{c}
-2.364\\
\left(0.417\right)
\end{array}$}  &  & {\footnotesize{}$\begin{array}{c}
-1.281\\
\left(0.043\right)
\end{array}$}  & {\footnotesize{}$\begin{array}{c}
-1.319\\
\left(0.045\right)
\end{array}$}  &  & {\footnotesize{}$\begin{array}{c}
-2.958\\
\left(0.597\right)
\end{array}$} & {\footnotesize{}$\begin{array}{c}
-3.842\\
\left(0.714\right)
\end{array}$} & {\footnotesize{}$\begin{array}{c}
-3.912\\
\left(0.567\right)
\end{array}$}\tabularnewline
{\footnotesize{}Cherry}  & {\footnotesize{}$\begin{array}{c}
-2.820\\
\left(18.229\right)
\end{array}$}  & {\footnotesize{}$\begin{array}{c}
-4.157\\
\left(0.472\right)
\end{array}$}  &  & {\footnotesize{}$\begin{array}{c}
-2.672\\
\left(2.073\right)
\end{array}$}  & {\footnotesize{}$\begin{array}{c}
-0.268\\
\left(3.573\right)
\end{array}$}  &  & {\footnotesize{}$\begin{array}{c}
-3.018\\
\left(0.312\right)
\end{array}$}  & {\footnotesize{}$\begin{array}{c}
-2.892\\
\left(0.351\right)
\end{array}$}  &  & {\footnotesize{}$\begin{array}{c}
-0.743\\
\left(5.991\right)
\end{array}$} & {\footnotesize{}$\begin{array}{c}
-0.779\\
\left(6.469\right)
\end{array}$} & {\footnotesize{}$\begin{array}{c}
-1.718\\
\left(5.142\right)
\end{array}$}\tabularnewline
{\footnotesize{}Coke}  & {\footnotesize{}$\begin{array}{c}
1.802\\
\left(0.509\right)
\end{array}$}  & {\footnotesize{}$\begin{array}{c}
2.631\\
\left(0.377\right)
\end{array}$}  &  & {\footnotesize{}$\begin{array}{c}
1.833\\
\left(0.186\right)
\end{array}$}  & {\footnotesize{}$\begin{array}{c}
0.252\\
\left(0.600\right)
\end{array}$}  &  & {\footnotesize{}$\begin{array}{c}
1.684\\
\left(0.106\right)
\end{array}$}  & {\footnotesize{}$\begin{array}{c}
1.633\\
\left(0.109\right)
\end{array}$}  &  & {\footnotesize{}$\begin{array}{c}
4.432\\
\left(1.233\right)
\end{array}$} & {\footnotesize{}$\begin{array}{c}
6.665\\
\left(1.249\right)
\end{array}$} & {\footnotesize{}$\begin{array}{c}
4.610\\
\left(0.993\right)
\end{array}$}\tabularnewline
{\footnotesize{}Pepsi}  & {\footnotesize{}$\begin{array}{c}
1.366\\
\left(0.081\right)
\end{array}$}  & {\footnotesize{}$\begin{array}{c}
1.456\\
\left(0.054\right)
\end{array}$}  &  & {\footnotesize{}$\begin{array}{c}
1.446\\
\left(0.082\right)
\end{array}$}  & {\footnotesize{}$\begin{array}{c}
1.275\\
\left(0.130\right)
\end{array}$}  &  & {\footnotesize{}$\begin{array}{c}
1.464\\
\left(0.044\right)
\end{array}$}  & {\footnotesize{}$\begin{array}{c}
1.471\\
\left(0.044\right)
\end{array}$}  &  & {\footnotesize{}$\begin{array}{c}
12.259\\
\left(2.377\right)
\end{array}$} & {\footnotesize{}$\begin{array}{c}
15.105\\
\left(2.376\right)
\end{array}$} & {\footnotesize{}$\begin{array}{c}
12.212\\
\left(1.889\right)
\end{array}$}\tabularnewline
{\footnotesize{}Promo}  & {\footnotesize{}-}  & {\footnotesize{}$\begin{array}{c}
0.294\\
\left(0.466\right)
\end{array}$}  &  & {\footnotesize{}-}  & {\footnotesize{}$\begin{array}{c}
-3.216\\
\left(1.264\right)
\end{array}$}  &  & {\footnotesize{}-}  & {\footnotesize{}$\begin{array}{c}
-0.331\\
\left(0.128\right)
\end{array}$}  &  & {\footnotesize{}-} & {\footnotesize{}-} & {\footnotesize{}-}\tabularnewline
 &  &  &  &  &  &  &  &  &  &  &  & \tabularnewline
{\footnotesize{}$D$}  & {\footnotesize{}$3667$}  & {\footnotesize{}$3667$}  &  & {\footnotesize{}$3667$}  & {\footnotesize{}$3667$}  &  & {\footnotesize{}$3667$}  & {\footnotesize{}$3667$}  &  & {\footnotesize{}$3667$}  & {\footnotesize{}$4069$}  & {\footnotesize{}$4069$}\tabularnewline
{\footnotesize{}$N$}  & {\footnotesize{}$5185$}  & {\footnotesize{}$5185$}  &  & {\footnotesize{}$5185$}  & {\footnotesize{}$5185$}  &  & {\footnotesize{}$5185$}  & {\footnotesize{}$5185$}  &  & {\footnotesize{}$3667$}  & {\footnotesize{}$4069$}  & {\footnotesize{}$4069$}\tabularnewline
\hline 
\end{tabular}\medskip{}
\par\end{centering}
{\footnotesize{}Note. (i) For columns Our Model, K/S and Our Model,
Probit, results from the Klein-Spady and Probit estimators in Table
\ref{tab:First-stage-Parameter-Estimates-2} were used during the
first-stage estimator, respectively. A pairwise differenced weighted
instrumental variable estimator was then used during the second stage.
For the Heckman Correction column, the Probit was used during the
first stage, and the Heckman's selection correction estimator with
the inverse Mills ratio as an additional regressor was used during
the second stage. (ii) $D$ is the number of non-zero market share
observations, and $N$ is the effective sample size. (iii) Asymptotic
standard error estimates appear in parentheses. (iv) The unit of bottle
size is liquid ounces. (v) Because Dominick's did not record the price
and cost when sales were zero, when estimating the $10^{-8}$ and
$10^{-4}$ columns, we used average prices and costs of the same product
with the same promotion statuses from other stores.}{\footnotesize\par}
\end{sidewaystable}

\newpage{}

\subsection{\label{subsec:Laundry-Detergent-Data}Laundry Detergent Demand for
Week 375}

We estimate demand for laundry detergent using the same data from
Dominick's as in Section \ref{sec:Empirical-Example}. We chose a
cross-section of week 375 randomly, which is the third week of November
1996. The product was defined by its UPC, and the market was defined
as the week-store pair. We compute market shares by loads. There are
two types of laundry detergents\textendash liquid and powder. We convert
the size of a canister by the following criteria. For liquid laundry
detergent, 1.6 ounces were counted as one load. For powder laundry
detergent, 2.3 ounces were counted as one load. Some powder detergents
used pounds instead of ounces as a unit of package size.\footnote{Arm \& Hammer powdered detergent.}
For such products, 0.08262 pounds was counted as one load. As the
density of powder detergents are approximately $0.65\mbox{g/c\ensuremath{m^{3}}}$
and $1\mbox{g/c\ensuremath{m^{3}}}=0.065198\mbox{lb/oz}$, one pound
of powdered detergent is approximately $23.6$ ounces. Market size
was calculated assuming that each consumer who visited the store consumed
6 loads of laundry detergent each week. Other details on the data
were similar to what we describe in Section \ref{sec:Empirical-Example}.

Table \ref{tab:First-stage-Parameter-Estimates-1-1} shows first-stage
estimates, and Table \ref{tab:Parameter-Estimates-1-1} shows those
for the second stage. We find the same pattern as in Table \ref{tab:Parameter-Estimates}
in Section \ref{sec:Empirical-Example}; even after instrumenting
for prices, the estimated demand curve was upward-sloping when consideration
set selection was not considered during estimation. 

\begin{table}[!tbph]
\caption{\label{tab:First-stage-Parameter-Estimates-1-1}First-stage Parameter
Estimates $\hat{\bm{\delta}}$ }

\begin{centering}
\begin{tabular}{ccc}
\hline 
 & {\footnotesize{}Probit}  & {\footnotesize{}Klein-Spady}\tabularnewline
{\footnotesize{}$\mathbf{w}_{j,t}$} & {\footnotesize{}(1)} & {\footnotesize{}(2)}\tabularnewline
\hline 
{\footnotesize{}Package Size}  & {\footnotesize{}$\begin{array}{c}
1\\
\left(0.091\right)
\end{array}$}  & {\footnotesize{}$\begin{array}{c}
1\\
\left(-\right)
\end{array}$}\tabularnewline
{\footnotesize{}Liquid}  & {\footnotesize{}$\begin{array}{c}
379.843\\
\left(16.031\right)
\end{array}$}  & {\footnotesize{}$\begin{array}{c}
320.578\\
\left(5.676\right)
\end{array}$}\tabularnewline
{\footnotesize{}Heavy duty / Concentrated / Double}  & {\footnotesize{}$\begin{array}{c}
114.862\\
\left(42.456\right)
\end{array}$}  & {\footnotesize{}$\begin{array}{c}
-18.295\\
\left(5.571\right)
\end{array}$}\tabularnewline
{\footnotesize{}Bleach}  & {\footnotesize{}$\begin{array}{c}
-6.144\\
\left(18.997\right)
\end{array}$}  & {\footnotesize{}$\begin{array}{c}
-0.279\\
\left(2.714\right)
\end{array}$}\tabularnewline
{\footnotesize{}Tide}  & {\footnotesize{}$\begin{array}{c}
230.570\\
\left(20.483\right)
\end{array}$}  & {\footnotesize{}$\begin{array}{c}
266.716\\
\left(5.676\right)
\end{array}$}\tabularnewline
{\footnotesize{}Wisk}  & {\footnotesize{}$\begin{array}{c}
31.895\\
\left(27.169\right)
\end{array}$}  & {\footnotesize{}$\begin{array}{c}
-94.849\\
\left(4.451\right)
\end{array}$}\tabularnewline
{\footnotesize{}Ajax / Arm\&Hammer / Surf / Purex}  & {\footnotesize{}$\begin{array}{c}
-240.998\\
\left(22.464\right)
\end{array}$}  & {\footnotesize{}$\begin{array}{c}
-152.440\\
\left(4.440\right)
\end{array}$}\tabularnewline
{\footnotesize{}\% Blacks and Hispanics}  & {\footnotesize{}$\begin{array}{c}
18.464\\
\left(57.478\right)
\end{array}$}  & {\footnotesize{}$\begin{array}{c}
-4.242\\
\left(8.147\right)
\end{array}$}\tabularnewline
{\footnotesize{}\% College Graduates}  & {\footnotesize{}$\begin{array}{c}
242.359\\
\left(91.294\right)
\end{array}$}  & {\footnotesize{}$\begin{array}{c}
1.831\\
\left(13.016\right)
\end{array}$}\tabularnewline
{\footnotesize{}Log Median Income}  & {\footnotesize{}$\begin{array}{c}
-9.221\\
\left(48.458\right)
\end{array}$}  & {\footnotesize{}$\begin{array}{c}
-2.217\\
\left(6.950\right)
\end{array}$}\tabularnewline
 &  & \tabularnewline
{\footnotesize{}$D$}  & {\footnotesize{}7177}  & {\footnotesize{}7177}\tabularnewline
{\footnotesize{}$N$}  & {\footnotesize{}14999}  & {\footnotesize{}14999}\tabularnewline
\hline 
\end{tabular}\medskip{}
\par\end{centering}
{\footnotesize{}Note. (i) $D$ is the number of non-zero market share
observations, and $N$ is the sample size. (ii) Asymptotic standard
error estimates appear in parentheses. (iii) The unit of bottle size
is liquid ounces. (iv) We normalized the coefficients of Package Size
variable to one. }{\footnotesize\par}
\end{table}

\begin{sidewaystable}[H]
\caption{\label{tab:Parameter-Estimates-1-1}Second-stage Parameter Estimates
$\left(\hat{\sigma},\hat{\bm{\beta}}\right)$}

\begin{centering}
\begin{tabular}{cccccccccc}
\hline 
 &  & \multicolumn{3}{c}{{\footnotesize{}Our Model}} &  & \multicolumn{4}{c}{{\footnotesize{}Logit Model}}\tabularnewline
\cline{2-5} \cline{7-10} 
 & {\footnotesize{}first-stage:}  & {\footnotesize{}K/S}  & {\footnotesize{}Probit}  & {\footnotesize{}Heckman}  &  & {\footnotesize{}Zero:}  & {\footnotesize{}Drop 0}  & {\footnotesize{}$10^{-8}$}  & {\footnotesize{}$10^{-4}$}\tabularnewline
{\footnotesize{}$\left(\phi_{j,t},\mathbf{x}_{j,t}\right)$ } &  & {\footnotesize{}(1)} & {\footnotesize{}(2)} & {\footnotesize{}(3)} &  &  & {\footnotesize{}(4)} & {\footnotesize{}(5)} & {\footnotesize{}(6)}\tabularnewline
\hline 
{\footnotesize{}Log Price ($-\sigma$)}  &  & {\footnotesize{}$\begin{array}{c}
-0.991\\
\left(0.069\right)
\end{array}$}  & {\footnotesize{}$\begin{array}{c}
-0.823\\
\left(0.194\right)
\end{array}$}  & {\footnotesize{}$\begin{array}{c}
-1.059\\
\left(0.044\right)
\end{array}$}  &  &  & {\footnotesize{}$\begin{array}{c}
6.455\\
\left(0.095\right)
\end{array}$}  & {\footnotesize{}$\begin{array}{c}
9.755\\
\left(0.138\right)
\end{array}$}  & {\footnotesize{}$\begin{array}{c}
6.371\\
\left(0.064\right)
\end{array}$}\tabularnewline
{\footnotesize{}Package Size}  &  & {\footnotesize{}$\begin{array}{c}
0.009\\
\left(0.003\right)
\end{array}$}  & {\footnotesize{}$\begin{array}{c}
0.004\\
\left(0.004\right)
\end{array}$}  & {\footnotesize{}$\begin{array}{c}
0.000\\
\left(0.001\right)
\end{array}$}  &  &  & {\footnotesize{}$\begin{array}{c}
0.030\\
\left(0.001\right)
\end{array}$}  & {\footnotesize{}$\begin{array}{c}
0.038\\
\left(0.001\right)
\end{array}$}  & {\footnotesize{}$\begin{array}{c}
0.026\\
\left(0.001\right)
\end{array}$}\tabularnewline
{\footnotesize{}Liquid}  &  & {\footnotesize{}$\begin{array}{c}
3.077\\
\left(0.985\right)
\end{array}$}  & {\footnotesize{}$\begin{array}{c}
1.585\\
\left(1.538\right)
\end{array}$}  & {\footnotesize{}$\begin{array}{c}
-0.032\\
\left(0.296\right)
\end{array}$}  &  &  & {\footnotesize{}$\begin{array}{c}
6.551\\
\left(0.153\right)
\end{array}$}  & {\footnotesize{}$\begin{array}{c}
9.990\\
\left(0.228\right)
\end{array}$}  & {\footnotesize{}$\begin{array}{c}
6.020\\
\left(0.106\right)
\end{array}$}\tabularnewline
{\footnotesize{}$\begin{array}{c}
\mbox{Heavy duty /}\\
\mbox{Concentrated / Double}
\end{array}$}  &  & {\footnotesize{}$\begin{array}{c}
-0.397\\
\left(0.104\right)
\end{array}$}  & {\footnotesize{}$\begin{array}{c}
0.318\\
\left(0.843\right)
\end{array}$}  & {\footnotesize{}$\begin{array}{c}
-0.435\\
\left(0.143\right)
\end{array}$}  &  &  & {\footnotesize{}$\begin{array}{c}
4.218\\
\left(0.249\right)
\end{array}$}  & {\footnotesize{}$\begin{array}{c}
7.803\\
\left(0.467\right)
\end{array}$}  & {\footnotesize{}$\begin{array}{c}
4.108\\
\left(0.217\right)
\end{array}$}\tabularnewline
{\footnotesize{}Bleach}  &  & {\footnotesize{}$\begin{array}{c}
-0.016\\
\left(0.039\right)
\end{array}$}  & {\footnotesize{}$\begin{array}{c}
-0.016\\
\left(0.053\right)
\end{array}$}  & {\footnotesize{}$\begin{array}{c}
-0.035\\
\left(0.054\right)
\end{array}$}  &  &  & {\footnotesize{}$\begin{array}{c}
0.575\\
\left(0.101\right)
\end{array}$}  & {\footnotesize{}$\begin{array}{c}
0.837\\
\left(0.172\right)
\end{array}$}  & {\footnotesize{}$\begin{array}{c}
0.720\\
\left(0.080\right)
\end{array}$}\tabularnewline
{\footnotesize{}Tide}  &  & {\footnotesize{}$\begin{array}{c}
3.510\\
\left(0.872\right)
\end{array}$}  & {\footnotesize{}$\begin{array}{c}
1.503\\
\left(0.793\right)
\end{array}$}  & {\footnotesize{}$\begin{array}{c}
0.826\\
\left(0.183\right)
\end{array}$}  &  &  & {\footnotesize{}$\begin{array}{c}
0.796\\
\left(0.099\right)
\end{array}$}  & {\footnotesize{}$\begin{array}{c}
2.055\\
\left(0.175\right)
\end{array}$}  & {\footnotesize{}$\begin{array}{c}
0.578\\
\left(0.081\right)
\end{array}$}\tabularnewline
{\footnotesize{}Wisk}  &  & {\footnotesize{}$\begin{array}{c}
-0.491\\
\left(0.257\right)
\end{array}$}  & {\footnotesize{}$\begin{array}{c}
0.038\\
\left(0.103\right)
\end{array}$}  & {\footnotesize{}$\begin{array}{c}
-0.033\\
\left(0.071\right)
\end{array}$}  &  &  & {\footnotesize{}$\begin{array}{c}
-0.613\\
\left(0.137\right)
\end{array}$}  & {\footnotesize{}$\begin{array}{c}
-1.118\\
\left(0.229\right)
\end{array}$}  & {\footnotesize{}$\begin{array}{c}
-0.742\\
\left(0.107\right)
\end{array}$}\tabularnewline
{\footnotesize{}$\begin{array}{c}
\mbox{Ajax / ArmHammer /}\\
\mbox{ Surf / Purex}
\end{array}$}  &  & {\footnotesize{}$\begin{array}{c}
-0.722\\
\left(0.437\right)
\end{array}$}  & {\footnotesize{}$\begin{array}{c}
-0.490\\
\left(0.806\right)
\end{array}$}  & {\footnotesize{}$\begin{array}{c}
0.323\\
\left(0.199\right)
\end{array}$}  &  &  & {\footnotesize{}$\begin{array}{c}
1.715\\
\left(0.133\right)
\end{array}$}  & {\footnotesize{}$\begin{array}{c}
3.180\\
\left(0.226\right)
\end{array}$}  & {\footnotesize{}$\begin{array}{c}
2.034\\
\left(0.105\right)
\end{array}$}\tabularnewline
 &  &  &  &  &  &  &  &  & \tabularnewline
{\footnotesize{}$D$}  &  & {\footnotesize{}7177}  & {\footnotesize{}7177}  & {\footnotesize{}7177}  &  &  & {\footnotesize{}7177}  & {\footnotesize{}10639}  & {\footnotesize{}10639}\tabularnewline
{\footnotesize{}$N$}  &  & {\footnotesize{}14999}  & {\footnotesize{}14999}  & {\footnotesize{}14999}  &  &  & {\footnotesize{}7177}  & {\footnotesize{}10639}  & {\footnotesize{}10639}\tabularnewline
\hline 
\end{tabular}\medskip{}
\par\end{centering}
{\footnotesize{}Note. (i) For columns Our Model, K/S and Our Model,
Probit, results from the Klein-Spady and Probit estimators in Table
\ref{tab:First-stage-Parameter-Estimates-1-1} were used for the first-stage
estimator, respectively. A pairwise differenced weighted instrumental
variable estimator was then used during the second stage. For the
Heckman Correction column, the Probit was used during the first stage,
and the Heckman's selection correction estimator with the inverse
Mills ratio as an additional regressor was used during the second
stage. (ii) $D$ is the number of non-zero market share observations,
and $N$ is the effective sample size. (iii) Asymptotic standard error
estimates appear in parentheses. (iv) The unit of package size was
converted to liquid ounces. (v) Because Dominick's did not record
the price and cost when sales were zero, when estimating the $10^{-8}$
and $10^{-4}$ columns, we used average prices and costs of the same
product with the same promotion statuses from other stores.}{\footnotesize\par}
\end{sidewaystable}

\newpage{}

\subsection{\label{subsec:Laundry-Detergent-Demand}Laundry Detergent Demand
for Week 398}

We estimate demand for laundry detergent using the same data from
Dominick's as in the previous subsection. We selected a cross-section
of week 398, which is the last week of April 1997. Other details on
data handling were the same as in the previous subsection.

\begin{table}[!tbph]
\caption{\label{tab:First-stage-Parameter-Estimates-1-1-1}First-stage Parameter
Estimates $\hat{\bm{\delta}}$ }

\begin{centering}
\begin{tabular}{ccc}
\hline 
 & {\footnotesize{}Probit}  & {\footnotesize{}Klein-Spady}\tabularnewline
{\footnotesize{}$\mathbf{w}_{j,t}$ } & {\footnotesize{}(1)} & {\footnotesize{}(2)}\tabularnewline
\hline 
{\footnotesize{}Package Size}  & {\footnotesize{}$\begin{array}{c}
1\\
\left(0.101\right)
\end{array}$}  & {\footnotesize{}$\begin{array}{c}
1\\
\left(-\right)
\end{array}$}\tabularnewline
{\footnotesize{}Liquid}  & {\footnotesize{}$\begin{array}{c}
345.336\\
\left(17.758\right)
\end{array}$}  & {\footnotesize{}$\begin{array}{c}
449.144\\
\left(6.124\right)
\end{array}$}\tabularnewline
{\footnotesize{}Heavy duty / Concentrated / Double}  & {\footnotesize{}$\begin{array}{c}
111.494\\
\left(48.905\right)
\end{array}$}  & {\footnotesize{}$\begin{array}{c}
123.172\\
\left(5.447\right)
\end{array}$}\tabularnewline
{\footnotesize{}Bleach}  & {\footnotesize{}$\begin{array}{c}
-1.811\\
\left(20.737\right)
\end{array}$}  & {\footnotesize{}$\begin{array}{c}
6.195\\
\left(2.008\right)
\end{array}$}\tabularnewline
{\footnotesize{}Tide}  & {\footnotesize{}$\begin{array}{c}
341.478\\
\left(23.082\right)
\end{array}$}  & {\footnotesize{}$\begin{array}{c}
387.581\\
\left(7.019\right)
\end{array}$}\tabularnewline
{\footnotesize{}Wisk}  & {\footnotesize{}$\begin{array}{c}
191.418\\
\left(31.135\right)
\end{array}$}  & {\footnotesize{}$\begin{array}{c}
242.408\\
\left(7.153\right)
\end{array}$}\tabularnewline
{\footnotesize{}Ajax / Arm\&Hammer / Surf / Purex}  & {\footnotesize{}$\begin{array}{c}
-94.144\\
\left(24.119\right)
\end{array}$}  & {\footnotesize{}$\begin{array}{c}
-143.853\\
\left(3.401\right)
\end{array}$}\tabularnewline
{\footnotesize{}\% Blacks and Hispanics}  & {\footnotesize{}$\begin{array}{c}
101.921\\
\left(63.318\right)
\end{array}$}  & {\footnotesize{}$\begin{array}{c}
-3.192\\
\left(6.717\right)
\end{array}$}\tabularnewline
{\footnotesize{}\% College Graduates}  & {\footnotesize{}$\begin{array}{c}
141.812\\
\left(101.556\right)
\end{array}$}  & {\footnotesize{}$\begin{array}{c}
-3.712\\
\left(11.025\right)
\end{array}$}\tabularnewline
{\footnotesize{}Log Median Income}  & {\footnotesize{}$\begin{array}{c}
21.230\\
\left(52.890\right)
\end{array}$}  & {\footnotesize{}$\begin{array}{c}
-0.033\\
\left(5.735\right)
\end{array}$}\tabularnewline
 &  & \tabularnewline
{\footnotesize{}$D$}  & {\footnotesize{}7177}  & {\footnotesize{}7177}\tabularnewline
{\footnotesize{}$N$}  & {\footnotesize{}14999}  & {\footnotesize{}14999}\tabularnewline
\hline 
\end{tabular}\medskip{}
\par\end{centering}
{\footnotesize{}Note. (i) $D$ is the number of non-zero market share
observations, and $N$ is the sample size. (ii) Asymptotic standard
error estimates appear in parentheses. (iii) The unit of bottle size
was liquid ounces. (iv) We normalized the coefficients of Package
Size variable to one. }{\footnotesize\par}
\end{table}

\begin{sidewaystable}[H]
\caption{\label{tab:Parameter-Estimates-1-1-1}Second-stage Parameter Estimates
$\left(\hat{\sigma},\hat{\bm{\beta}}\right)$}

\begin{centering}
\begin{tabular}{cccccccccc}
\hline 
 &  & \multicolumn{3}{c}{{\footnotesize{}Our Model}} &  & \multicolumn{4}{c}{{\footnotesize{}Logit Model}}\tabularnewline
\cline{2-5} \cline{7-10} 
 & {\footnotesize{}first-stage:}  & {\footnotesize{}K/S}  & {\footnotesize{}Probit}  & {\footnotesize{}Heckman}  &  & {\footnotesize{}Zero:}  & {\footnotesize{}Drop 0}  & {\footnotesize{}$10^{-8}$}  & {\footnotesize{}$10^{-4}$}\tabularnewline
{\footnotesize{}$\left(\phi_{j,t},\mathbf{x}_{j,t}\right)$ } &  & {\footnotesize{}(1)} & {\footnotesize{}(2)} & {\footnotesize{}(3)} &  &  & {\footnotesize{}(4)} & {\footnotesize{}(5)} & {\footnotesize{}(6)}\tabularnewline
\hline 
{\footnotesize{}Log Price ($-\sigma$)}  &  & {\footnotesize{}$\begin{array}{c}
-0.664\\
\left(0.080\right)
\end{array}$}  & {\footnotesize{}$\begin{array}{c}
-0.779\\
\left(0.135\right)
\end{array}$}  & {\footnotesize{}$\begin{array}{c}
-0.938\\
\left(0.051\right)
\end{array}$}  &  &  & {\footnotesize{}$\begin{array}{c}
3.287\\
\left(0.036\right)
\end{array}$}  & {\footnotesize{}$\begin{array}{c}
9.632\\
\left(0.157\right)
\end{array}$}  & {\footnotesize{}$\begin{array}{c}
6.617\\
\left(0.077\right)
\end{array}$} \tabularnewline
{\footnotesize{}Package Size}  &  & {\footnotesize{}$\begin{array}{c}
0.012\\
\left(0.001\right)
\end{array}$}  & {\footnotesize{}$\begin{array}{c}
0.007\\
\left(0.002\right)
\end{array}$}  & {\footnotesize{}$\begin{array}{c}
0.000\\
\left(0.001\right)
\end{array}$}  &  &  & {\footnotesize{}$\begin{array}{c}
0.009\\
\left(0.000\right)
\end{array}$} & {\footnotesize{}$\begin{array}{c}
0.037\\
\left(0.001\right)
\end{array}$} & {\footnotesize{}$\begin{array}{c}
0.028\\
\left(0.001\right)
\end{array}$}\tabularnewline
{\footnotesize{}Liquid}  &  & {\footnotesize{}$\begin{array}{c}
6.189\\
\left(0.629\right)
\end{array}$}  & {\footnotesize{}$\begin{array}{c}
2.554\\
\left(1.074\right)
\end{array}$}  & {\footnotesize{}$\begin{array}{c}
-0.070\\
\left(0.390\right)
\end{array}$}  &  &  & {\footnotesize{}$\begin{array}{c}
1.861\\
\left(0.065\right)
\end{array}$}  & {\footnotesize{}$\begin{array}{c}
9.688\\
\left(0.254\right)
\end{array}$}  & {\footnotesize{}$\begin{array}{c}
6.287\\
\left(0.124\right)
\end{array}$} \tabularnewline
{\footnotesize{}$\begin{array}{c}
\mbox{Heavy duty /}\\
\mbox{Concentrated / Double}
\end{array}$}  &  & {\footnotesize{}$\begin{array}{c}
0.660\\
\left(0.198\right)
\end{array}$}  & {\footnotesize{}$\begin{array}{c}
-0.156\\
\left(0.246\right)
\end{array}$}  & {\footnotesize{}$\begin{array}{c}
-0.699\\
\left(0.207\right)
\end{array}$}  &  &  & {\footnotesize{}$\begin{array}{c}
0.780\\
\left(0.160\right)
\end{array}$} & {\footnotesize{}$\begin{array}{c}
0.087\\
\left(0.488\right)
\end{array}$} & {\footnotesize{}$\begin{array}{c}
4.105\\
\left(0.238\right)
\end{array}$}\tabularnewline
{\footnotesize{}Bleach}  &  & {\footnotesize{}$\begin{array}{c}
0.084\\
\left(0.042\right)
\end{array}$}  & {\footnotesize{}$\begin{array}{c}
0.044\\
\left(0.083\right)
\end{array}$}  & {\footnotesize{}$\begin{array}{c}
0.040\\
\left(0.071\right)
\end{array}$}  &  &  & {\footnotesize{}$\begin{array}{c}
-0.066\\
\left(0.063\right)
\end{array}$}  & {\footnotesize{}$\begin{array}{c}
1.388\\
\left(0.189\right)
\end{array}$}  & {\footnotesize{}$\begin{array}{c}
0.803\\
\left(0.092\right)
\end{array}$} \tabularnewline
{\footnotesize{}Tide}  &  & {\footnotesize{}$\begin{array}{c}
5.678\\
\left(0.538\right)
\end{array}$}  & {\footnotesize{}$\begin{array}{c}
2.620\\
\left(1.045\right)
\end{array}$}  & {\footnotesize{}$\begin{array}{c}
-0.028\\
\left(0.371\right)
\end{array}$}  &  &  & {\footnotesize{}$\begin{array}{c}
-0.262\\
\left(0.062\right)
\end{array}$} & {\footnotesize{}$\begin{array}{c}
1.579\\
\left(0.190\right)
\end{array}$} & {\footnotesize{}$\begin{array}{c}
-0.171\\
\left(0.092\right)
\end{array}$}\tabularnewline
{\footnotesize{}Wisk}  &  & {\footnotesize{}$\begin{array}{c}
3.657\\
\left(0.441\right)
\end{array}$}  & {\footnotesize{}$\begin{array}{c}
1.186\\
\left(0.562\right)
\end{array}$}  & {\footnotesize{}$\begin{array}{c}
-0.305\\
\left(0.231\right)
\end{array}$}  &  &  & {\footnotesize{}$\begin{array}{c}
-1.196\\
\left(0.088\right)
\end{array}$}  & {\footnotesize{}$\begin{array}{c}
1.025\\
\left(0.258\right)
\end{array}$}  & {\footnotesize{}$\begin{array}{c}
0.097\\
\left(0.126\right)
\end{array}$} \tabularnewline
{\footnotesize{}$\begin{array}{c}
\mbox{Ajax / ArmHammer /}\\
\mbox{ Surf / Purex}
\end{array}$}  &  & {\footnotesize{}$\begin{array}{c}
-1.003\\
\left(0.178\right)
\end{array}$}  & {\footnotesize{}$\begin{array}{c}
-0.105\\
\left(0.332\right)
\end{array}$}  & {\footnotesize{}$\begin{array}{c}
0.191\\
\left(0.144\right)
\end{array}$}  &  &  & {\footnotesize{}$\begin{array}{c}
0.601\\
\left(0.081\right)
\end{array}$} & {\footnotesize{}$\begin{array}{c}
2.719\\
\left(0.227\right)
\end{array}$} & {\footnotesize{}$\begin{array}{c}
1.681\\
\left(0.111\right)
\end{array}$}\tabularnewline
 &  &  &  &  &  &  &  &  & \tabularnewline
{\footnotesize{}$D$}  &  & {\footnotesize{}6339}  & {\footnotesize{}6339}  & {\footnotesize{}6339}  &  &  & {\footnotesize{}6339}  & {\footnotesize{}8587} & {\footnotesize{}8587}\tabularnewline
{\footnotesize{}$N$}  &  & {\footnotesize{}13625}  & {\footnotesize{}13625}  & {\footnotesize{}13625}  &  &  & {\footnotesize{}13625}  & {\footnotesize{}8587}  & {\footnotesize{}8587}\tabularnewline
\hline 
\end{tabular}\medskip{}
\par\end{centering}
{\footnotesize{}Note. (i) For columns Our Model, K/S and Our Model,
Probit, results from the Klein-Spady and Probit estimators in Table
\ref{tab:First-stage-Parameter-Estimates-1-1-1} were used for the
first-stage estimator, respectively. A pairwise differenced weighted
instrumental variable estimator was then used during the second stage.
For the Heckman Correction column, the Probit was used during the
first stage, and the Heckman's selection correction estimator with
the inverse Mills ratio as an additional regressor was used during
the second stage. (ii) $D$ is the number of non-zero market share
observations, and $N$ is the effective sample size. (iii) Asymptotic
standard error estimates appear in parentheses. (iv) The unit of package
size was converted to liquid ounces. (v) Because Dominick's did not
record the price and cost when sales were zero, when estimating the
$10^{-8}$ and $10^{-4}$ columns, we used average prices and costs
of the same product with the same promotion statuses from other stores.}{\footnotesize\par}
\end{sidewaystable}

\end{document}